# Impact probability computation of near-Earth objects using Monte Carlo line sampling and subset simulation

**Matteo Romano**[1] · **Matteo Losacco**[1] · **Camilla Colombo**[1] · **Pierluigi Di Lizia**[1]

**Abstract**
This work introduces two Monte Carlo (MC)-based sampling methods, known as line sampling and subset simulation, to improve the performance of standard MC analyses in the context of asteroid impact risk assessment. Both techniques sample the initial uncertainty region in different ways, with the result of either providing a more accurate estimate of the impact probability or reducing the number of required samples during the simulation with respect to standard MC techniques. The two methods are first described and then applied to some test cases, providing evidence of the increased accuracy or the reduced computational burden with respect to a standard MC simulation. Finally, a sensitivity analysis is carried out to show how parameter setting affects the accuracy of the results and the numerical efficiency of the two methods.

## 1 Introduction

Earth is subject to frequent impacts by small meteoroids and asteroids (Harris and D'Abramo 2015). Asteroids orbit the Sun along orbits that can allow them to enter the Earth's neighbourhood (near-Earth asteroids, NEAs), leading to periodic close approaches with our planet with the possibility of impacts on the ground and risks for human activity in space. In parallel, during interplanetary missions, launcher stages and inactive spacecraft are often left into orbits that may come back to the Earth or reach other celestial bodies, with the risk of impacting

✉ Matteo Romano
matteo1.romano@polimi.it

Matteo Losacco
matteo.losacco@polimi.it

Camilla Colombo
camilla.colombo@polimi.it

Pierluigi Di Lizia
pierluigi.dilizia@polimi.it

[1] Department of Aerospace Science and Technology, Politecnico di Milano, via La Masa 34, 20156 Milan, Italy

and contaminating them. For this reason, planetary protection policies set specific requirements to avoid the contamination of celestial bodies due to man-made debris in interplanetary missions, with time periods under study that generally span up to 100 years (Kminek 2012). The estimation and propagation of the orbital state of these objects is therefore of paramount importance.

Current approaches for robust detection and prediction of planetary encounters mainly refer to linearised models or full nonlinear orbital sampling. The application of linear methods in the impact plane was introduced by Chodas (1993), whereas the introduction of the Monte Carlo technique to this problem was developed by Yeomans and Chodas (1994) and Chodas and Yeomans (1999), and it is based on the sampling of the linear six-dimensional confidence region of the initial conditions, whose integration over the time interval of investigation uses fully nonlinear equations (Milani et al. 2002). Milani (1999), Milani et al. (2000a, b) and Milani et al. (2005) applied the multiple solutions approach to sample the central line of variations (LOV) of the (potentially nonlinear) confidence region. This method currently represents the standard approach for impact monitoring.

The preferred approach depends on the uncertainty in the estimated orbit, the investigated time window, and the dynamics between the observation epoch and the epoch of the expected impact. As described in Farnocchia et al. (2015), linear methods are preferred when linear approximations are reliable for both the orbit determination and uncertainty propagation. When these assumptions are not valid, more computationally intensive techniques are used: among these, Monte Carlo methods are the most accurate but also the most computationally intensive, whereas the LOV method guarantees computing times 3–4 orders of magnitude lower than those required in MC simulations, which makes this method the most efficient approach for impact monitoring for the vast majority of the scenarios. Nevertheless, there are some cases in which the LOV method does not guarantee the same level of accuracy of a standard MC approach. A first case occurs when the observed arc of the investigated object is very short, i.e. a couple of days or less (Milani et al. 2005). In this case, the confidence region is wide in two directions and the unidimensional sampling may not be suitable. What happens is that different LOVs, computed with different coordinates, provide independent sampling and may provide different results. That is, the choice of VAs along the 1D LOV of the initial conditions may not correctly capture the suite of different future dynamical paths (Milani et al. 2002). A second critical scenario occurs when the LOV is stretched and folded by multiple close planetary encounters (Farnocchia et al. 2015). In all these cases, standard MC offers more reliable results and it is generally used, despite unavoidable drawbacks in terms of computational time. Thus, the availability of alternative MC methods employing a lower number of samples would be positive.

In this paper, we present two advanced Monte Carlo methods, known as line sampling (Schuëller et al. 2004) and subset simulation (Au and Beck 2001), and we show their possible application to NEA impact probability computation. The line sampling method probes the impact region of the uncertainty domain by using lines instead of points. The impact probability is then estimated via analytical integration along such lines, resulting in a more accurate estimate. The subset simulation method computes the impact probability as the product of larger conditional probabilities. The method progressively identifies intermediate conditional levels moving towards the impact event, reducing the overall number of samples required for the estimation. Originally developed to estimate failure probabilities of structural systems, the two methods have been recently proposed in combination with differential algebra (DA) for orbital conjunctions analysis by Morselli et al. (2015), whereas a preliminary analysis of the suitability of the use of the SS method for NEA impact probability computation was presented by Losacco et al. (2018) in their work. This paper starts from the latter and extends the analysis to the LS method, aiming at proposing a general scheme for the use of both methods for impact monitoring and offering a detailed analysis of their performance against different test cases and of their sensitivity to the available degrees of freedom. The paper is organised as follows. Sections 2 and 3 show a detailed theoretical descriptions of the two methods. Then, the results of three different test cases are presented in Sect. 4, and a comparison with the performance of standard MC is shown. Finally, the sensitivity of both methods to parameter setting is investigated in Sect. 5.

## 2 Line sampling

The line sampling (LS) method is a Monte Carlo-based approach for the estimation of small probabilities. Originally developed for the reliability analysis of failure in complex structural systems (Schuëller et al. 2004), and recently presented in combination with DA for orbital conjunctions analysis by Morselli et al. (2015), the method is here adapted to the estimation of the impact probability of small objects with major celestial bodies. The method estimates the impact probability via analytical evaluation by identifying the boundaries of the impact regions within the uncertainty domain, i.e. the set of initial conditions that lead to an impact within the given time. This result is obtained by considering several one-dimensional problems across the uncertainty domain. The analytical evaluations are carried out along lines following a reference direction, which is determined so that it points towards the impact region of the domain. If this direction is properly chosen, the method can considerably reduce the number of required propagations with respect to a standard MC.

The application of the LS method for impact probability computation requires an a priori approximate knowledge of the epoch at which the close approach may take place. In the current implementation of the method, the identification of the events is done with a preliminary survey, here defined as phase 0 of the LS method. Starting from the knowledge of the available state estimate of a newly discovered object, the phase 0 consists in performing a MC survey for a selected time frame (e.g. 100 years) with a relatively low number of samples. This survey provides a list of epochs of possible close approaches, which are identified by examining the planetocentric distance of all the propagated samples with respect to the celestial body of interest, and identifying all the epochs for which the minimum distance is lower than an imposed threshold. This preliminary phase provides a rough census of the epochs of possible close approaches but no impact probability associated with any of these events, which are then treated independently by the LS method. Once defined the set of event epochs, the method adopts the minimum geocentric distance as performance index and requires the identification of a proper time interval $\Delta t_e$ around each identified event $t_e$ where the index is computed. The time interval is arbitrarily selected as a window of 200 days centred at the estimated event epoch.

After identifying all possible epochs and selecting one target event $t_e$, the LS method has four steps: (1) the mapping of random samples from the physical coordinate space into a normalised standard space; (2) the determination of the reference direction $\boldsymbol{\alpha}$; (3) the probing of the impact region along the lines following the reference direction; (4) the estimation of the impact probability.

### 2.1 Mapping onto the standard normal space

The first step of the LS procedure is the definition of the mapping onto the standard normal space, which involves all the random vectors $\boldsymbol{x} \in R^n$ of physical coordinates (position and velocity) that are drawn from the nominal uncertainty distribution during the LS process. This transformation grants efficiency to the method, especially for problems with high dimensionality, as each component $\theta_j,\ j = 1, \ldots, n$ of the new parameter vector $\boldsymbol{\theta} \in R^n$, to which $\boldsymbol{x}$ is mapped, is associated with a standard normal distribution. The joint probability density function (pdf) of these random parameters is

$$\phi(\boldsymbol{\theta}) = \prod_{j=1}^{n} \phi_j(\theta_j), \tag{1}$$

where $\phi_j$ denotes the unit Gaussian pdf associated with the $j$-th component of $\boldsymbol{\theta}$ (Zio and Pedroni 2009):

$$\phi_j(\theta_j) = \frac{1}{\sqrt{2\pi}} \exp\left(-\frac{\theta_j^2}{2}\right), \quad j = 1, \ldots, n. \tag{2}$$

This enables a simplification of the computation of the probability later in the procedure, as it reduces the problem to a series of one-dimensional analytical evaluations. The direct and the inverse transformations, from the physical domain to the standardised one and vice versa, preserve the joint cumulative distribution function (CDF) between the two coordinate spaces, and they are defined as:

$$\Phi(\boldsymbol{\theta}) = F(\boldsymbol{x}), \tag{3}$$

$$\boldsymbol{\theta} = \Phi^{-1}[F(\boldsymbol{x})], \tag{4}$$

$$\boldsymbol{x} = F^{-1}[\Phi(\boldsymbol{\theta})], \tag{5}$$

with $\Phi$ and $F$ being the CDF of the unit Gaussian distribution and the input uncertainty distribution of the problem, respectively. Following the definition of the pdf $\phi$, the joint CDF $\Phi$ is

$$\begin{aligned} \Phi(\boldsymbol{\theta}) &= \int_{-\infty}^{\boldsymbol{\theta}} \phi(\boldsymbol{u}) \mathrm{d}\mathbf{u} = \prod_{j=1}^{n} \Phi_j(\theta_j) \\ \text{with} \quad \Phi_j(\theta_j) &= \frac{1}{2}\left[1 + \operatorname{erf}\left(\frac{\theta_j}{\sqrt{2}}\right)\right], \quad j = 1, \ldots, n, \end{aligned} \tag{6}$$

where $\operatorname{erf}(x) = \frac{2}{\sqrt{\pi}} \int_0^x \exp(-u^2) \mathrm{d}u$ is the error function.

The Rosenblatt transformation is applied in this work (Rosenblatt 1952), since, for Gaussian-distributed uncertainty parameters, both the direct and the inverse transformations (respectively, Eqs. 4 and 5) become linear (Zio and Pedroni 2009; Rosenblatt 1952). This choice was made since, in the case under study, the distribution of the initial conditions (position and velocity) is assumed to be Gaussian.

### 2.2 Determination of the reference direction

The reference direction $\boldsymbol{\alpha}$ can be determined in different ways (Zio and Pedroni 2009; Zio 2013). In this work, it is determined as the direction of a normalised "center of mass" of the impact region. This region is approximated by applying the Metropolis–Hastings algorithm (Metropolis et al. 1953; Hastings 1970) to generate a Monte Carlo Markov chain (MCMC) lying entirely in the impact subdomain starting from an initial condition within it. MCMC simulation is a method for generating conditional samples according to any given probability distribution described by the pdf $p(\mathbf{x})$. The algorithm to generate a sequence of $N_S$ samples from a given sample $\mathbf{x}_u$ drawn from the distribution $p(\mathbf{x})$ is briefly explained in Au and Beck (2001):

1. generate the sample $\boldsymbol{\xi}$ by randomly sampling a user-defined "proposal" pdf $p^*(\mathbf{x}_u)$: in this work, the proposal pdf used to build the MCMC for the LS method is obtained by applying a scaling of 1/10 to the initial uncertainty distributions, which allows the method to draw samples in the vicinity of the impact region;
2. compute the ratio $r = p(\boldsymbol{\xi})/p(\mathbf{x}_u)$;
3. set $\tilde{\mathbf{x}} = \boldsymbol{\xi}$ with probability $\min(1, r)$ and $\tilde{\mathbf{x}} = \mathbf{x}_u$ with the probability $1 - \min(1, r)$, where $\tilde{\mathbf{x}}$ is the candidate for the next element of the chain;
4. check whether the candidate $\tilde{\mathbf{x}}$ lies in the region of interest $I$ or not: if $\tilde{\mathbf{x}} \in I$, accept it as the next sample $\mathbf{x}_{u+1} = \tilde{\mathbf{x}}$; else, reject it and take the current sample as the next sample $\mathbf{x}_{u+1} = \mathbf{x}_u$.

In the previous formulation, the region of interest $I$ is the impact region for the selected event $t_e$, which is characterised by samples with planetocentric distance lower than the planet radius as computed in the time interval $\Delta t_e$.

The starting condition for the MCMC can be found with different approaches, such as optimisation processes or MC sampling. In the proposed method, the starting condition is computed with an optimisation process that exploits the MATLAB *fmincon* function. The process starts from the nominal initial conditions and aims at finding an initial state minimising the planetocentric distance of the object from the target body in the time interval $\Delta t_e$. The result is a solution which is close or inside the impact region.

The reference direction $\boldsymbol{\alpha}$ is then computed in the standard normal space as

$$\boldsymbol{\alpha} = \frac{\sum_{u=0}^{N_S} \boldsymbol{\theta}^u / N_S}{\| \sum_{u=0}^{N_S} \boldsymbol{\theta}^u / N_S \|}, \tag{7}$$

where $\boldsymbol{\theta}^u, \ u = 1, \ldots, N_S$ are the points of the Markov chain made of $N_{MCMC}$ samples found in the impact region of the uncertainty domain, converted from the physical space into the standard normal space. The simulations performed for the Markov chain require additional computational effort with respect to standard MC methods. Nevertheless, this option provides a good coverage of the impact region and a resulting better accuracy of the final probability estimate.

### 2.3 Line sampling

After determining the reference sampling direction, $N_T$ initial conditions $\boldsymbol{x}^k, \ k = 1, \ldots, N_T$ are randomly drawn from the nominal uncertainty distribution and then mapped to standard normal coordinates as $\boldsymbol{\theta}^k$ using the transformation in Eq. (4). For each sample in the standard normal space, a line starting from $\boldsymbol{\theta}^k, \ k = 1, \ldots, N_T$ and parallel to $\boldsymbol{\alpha}$ is defined according to the parameter $c^k$, such that

$$\tilde{\boldsymbol{\theta}}^k = c^k \boldsymbol{\alpha} + \boldsymbol{\theta}^{k,\perp}, \tag{8}$$

$$\boldsymbol{\theta}^{k,\perp} = \boldsymbol{\theta}^k - \langle \boldsymbol{\alpha}, \boldsymbol{\theta}^k \rangle \boldsymbol{\alpha}, \tag{9}$$

**Fig. 1** Scheme of the iterative procedure used to sample each line in the standard normal coordinate space. The impact region is represented with a single border highlighted as a red line

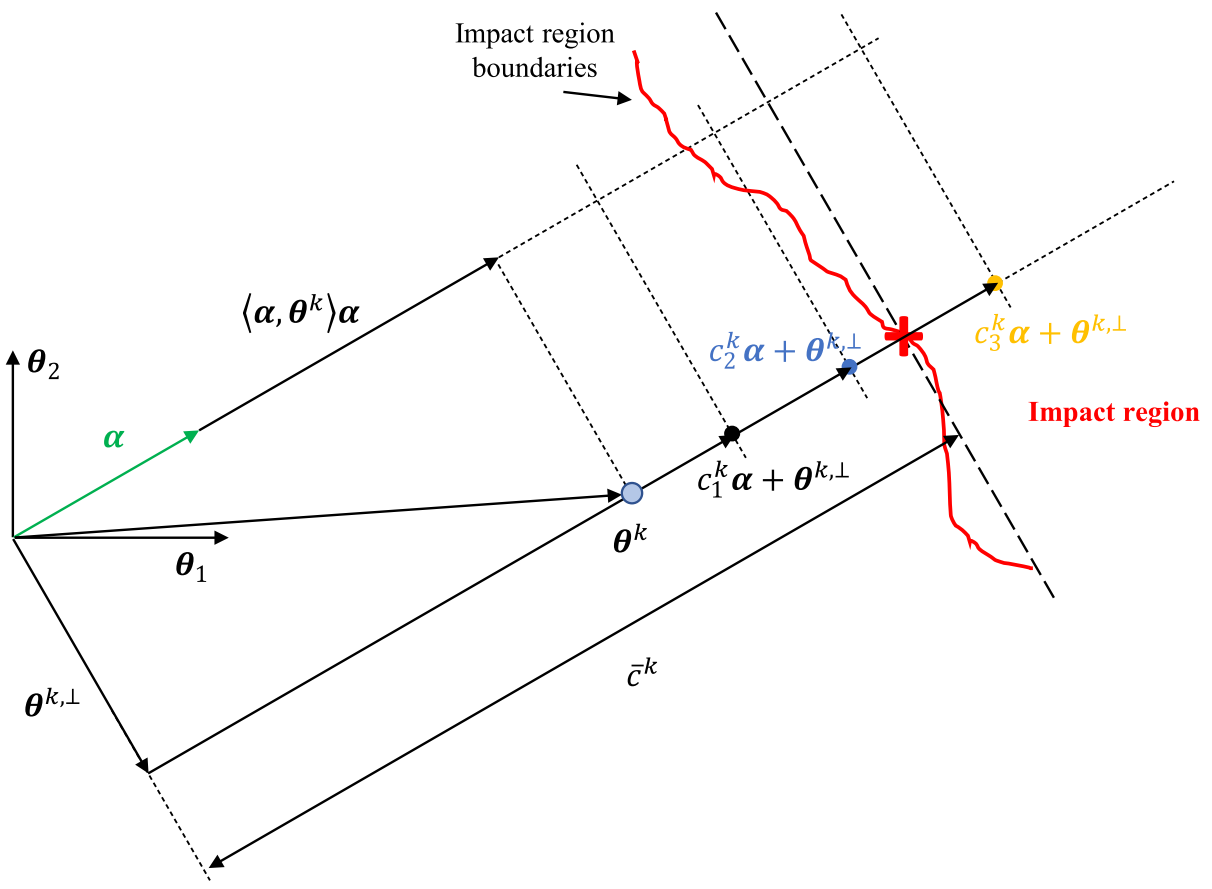

where $\langle\boldsymbol{\alpha}, \boldsymbol{\theta}^k\rangle$ is the scalar product between $\boldsymbol{\alpha}$ and $\boldsymbol{\theta}^k$. In this way, the problem is reduced to a series of one-dimensional evaluations associated with each sample, with $c^k$ being normally distributed in the standard space.

The standard domain is then explored along each line by iteratively evaluating a performance function $Y(c)$ to identify the values of $c^k$ corresponding to the (possible) intersections between the line and the impact region, as displayed in Fig. 1. The performance function considered in this work is defined as

$$Y(c) = d_{\Delta t_e}/R_P - 1, \tag{10}$$

where $d_{\Delta t_e}$ is the minimum distance from the celestial body of interest (e.g. the Earth) computed in $\Delta t_e$, and $R_P$ is the planet radius, making $Y(c)$ the non-dimensional minimum distance from the planet's surface. The minimum distance is computed numerically both in case the sphere of influence is crossed or not, so that $Y(c)$ can always be defined. According to this definition, it follows that

$$Y(c) \begin{cases} < 0 \rightarrow \text{Impact} \\ = 0 \rightarrow \text{Limit state} \\ > 0 \rightarrow \text{No impact} \end{cases}. \tag{11}$$

In this work, it is assumed that, due to the nature of the problem under analysis (single event within a given time interval), a maximum of two intersections between each line and the impact region can be found, meaning that, for each standard normal random sample $\boldsymbol{\theta}^k$, two values of $c^k$ at most exist such that the performance function is equal to zero: $Y(\bar{c}_1^k) = 0$ and $Y(\bar{c}_2^k) = 0$. The two possible solutions $(\bar{c}_1^k, \bar{c}_2^k)$ are identified with an iterative process, which implies some extra evaluations for each sample $\boldsymbol{\theta}^k$, $k = 1, \ldots, N_T$ with respect a standard MC simulation. The method adopted here makes use of Newton iterations, which require the knowledge of the derivative

$$\frac{\mathrm{d}Y(c^k)}{\mathrm{d}c^k} = \frac{\partial Y}{\partial d_{\Delta t_e}} \frac{\partial d_{\Delta t_e}}{\partial \boldsymbol{x}_f} \frac{\partial \boldsymbol{x}_f}{\partial \boldsymbol{x}_0} \frac{\partial \boldsymbol{x}_0}{\partial \boldsymbol{\theta}^k} \frac{\partial \boldsymbol{\theta}^k}{\partial c^k}, \tag{12}$$

where $\boldsymbol{x}_0$ and $\boldsymbol{x}_f$ indicate the initial and the final state of the propagation. While computing the partial derivatives of the various transformations from the parameter $c^k$ to the initial state (fourth and fifth terms) and from the final state to the minimum distance (first and second terms) is relatively simple, obtaining the partial derivatives of the final state with respect to the initial state requires the propagation of the state transition matrix, which increases the size of the system by six times, slows down the simulation and makes the implementation more complex. Since the state transition matrix is not necessary for the scope of this work, using two propagations to approximate the derivative numerically is a simpler choice in terms of implementation and computational cost. Moreover, the continuity and smoothness of $Y(c)$ in the vicinity of the impact region are granted under the hypothesis that the selected time interval is narrow enough to contain only a single close approach event to be analysed. When considering larger intervals, the function may show jump discontinuities due to the crossing of the sampling lines into a new region corresponding to a different close approach. However, under the hypotheses already considered, such cases do not occur during the analysis.

The derivative is here approximated numerically using the first-order forward scheme

$$\frac{\mathrm{d}Y(c^k)}{\mathrm{d}c^k} = \frac{Y(c^k + \Delta c) - Y(c^k)}{\Delta c}, \tag{13}$$

where the values of $Y$ are obtained propagating the initial states corresponding to the values $c^k$ and $c^k + \Delta c$, where $\Delta c$ is an arbitrarily small increment.

The assumption of two possible intersections per sampling line is generally valid when the impact region extends across the uncertainty domain and can be approximated as a flat or slightly curved surface. As a result, two outcomes are possible for each drawn sample $\boldsymbol{\theta}^k$: if two intersections are identified, the line crosses the impact region, and is thus considered for the impact probability computation; if no intersection is found, the line is assumed not to cross the event region, and therefore, it does not contribute to the impact probability estimation.

## 2.4 Estimation of the impact probability

Once the values $(\bar{c}_1^k, \bar{c}_2^k)$ are eventually known for all the sampling lines, the unit Gaussian CDF provides each random initial condition $\boldsymbol{\theta}^k$ with the conditional impact probability $\hat{P}^k(I)$, where

$$\hat{P}^k(I) = \hat{P}[\bar{c}_1^k < N(0,1) < \bar{c}_2^k] = \Phi(\bar{c}_2^k) - \Phi(\bar{c}_1^k) \tag{14}$$

if the two intersections exist, and $\hat{P}^k(I) = 0$ if no intersection is found. The total probability of the event $\hat{P}(I)$ and the associated variance $\hat{\sigma}^2(\hat{P}(I))$ are then approximated as

$$\hat{P}(I) = \frac{1}{N_T} \sum_{j=1}^{N_T} \hat{P}^k(I), \tag{15}$$

$$\hat{\sigma}^2(\hat{P}(I)) = \frac{1}{N_T(N_T - 1)} \sum_{j=1}^{N_T} (\hat{P}^k(I) - \hat{P}(I))^2. \tag{16}$$

The described procedure is repeated for all the identified events $t_e$. The completeness in the identification of the potential impacts and the accuracy in their characterisation depend on the parameters selected for the approach. The identification of the potential impacts is done during the phase 0. The selection of the threshold distance for the definition of the potential impacts determines the number of considered target epochs, and also the completeness of the analysis. In general, to avoid missing approaches on the way, a relatively large threshold is imposed. This may result into an overestimated number of events that are later analysed by the LS method. Then, the accuracy in the characterisation of each event is governed by the control parameters of the method. Among these, the computation of the reference direction $\boldsymbol{\alpha}$ and the identification of the boundaries of the impact region represent the most critical aspects of the method. An analysis of their role is offered in Sect. 5.

## 3 Subset simulation

The subset simulation (SS) method is a Monte Carlo method based on the principle of computing small event probabilities as the product of larger conditional probabilities (Au and Beck 2001; Cadini et al. 2012; Zuev et al. 2012). The method was originally developed for the identification of structural failures, but has been extended to different research areas, including the assessment of collision probability among resident space objects (see Morselli et al. 2015). Given a target event $I$, i.e. an event whose small probability is to be computed, let $I_1 \supset I_2 \supset \cdots \supset I_n = I$ be a sequence of intermediate events, so that $I_k = \cap_{i=1}^{k} I_i$. Given a sequence of conditional probabilities, the target event probability can be written as

$$P(I_n) = P(I_1) \prod_{i=1}^{n-1} P(I_{i+1}|I_i), \tag{17}$$

where $P(I_{i+1}|I_i)$ represents the probability of $I_{i+1}$ conditional to $I_i$. In the approach presented in this paper, the event is identified with a planetary collision, and the performance index used in the analysis is the minimum planetocentric distance of the propagated sample.

The SS method aims at identifying and investigating possible impact events independently. The identification of the events is done with a preliminary survey, here defined as run 0 of the SS method, which is very similar to the phase 0 of the LS method. Starting from the knowledge of the available state estimate of a newly discovered object, the run 0 consists in performing a Monte Carlo survey for a selected time frame. This survey provides a list of epochs of possible close approaches, which represent the input for the later stages of the SS method, and a set of $N$ samples at the so-called conditional level 0 (CL0). The method is then run to estimate the impact probability for each single identified event.

Similarly to the LS method, the element that distinguishes one event from another is the time interval at which the minimum distance is computed. Specifically, once a single target event and the associated epoch $t_e$ have been identified, the time interval where to compute the minimum geocentric distance $d_{\Delta t_e}$ for each drawn sample is defined as $\Delta t_e = \{t_{e,1}, t_{e,2}\} = \{t_e - 100\text{d}, t_e + 100\text{d}\}$. Then, after computing $d_{\Delta t_e}$ and sorting all the results, a first event region $I_1$ is identified, and an MCMC Metropolis–Hastings algorithm is used to generate conditional samples in the new region. At this stage, another intermediate region $I_2$ is then located, and other samples are generated. The procedure is repeated until the impact region is identified. An illustration of the SS method is given in Fig. 2.

In this work, the intermediate event regions are identified by assuming a fixed value of conditional probability $P(I_{i+1}|I_i) = p_0$. The identification of each conditional level is affected by this value and changes accordingly step by step, as explained hereafter. Following the general description offered in Morselli et al. (2015), the resulting SS algorithm goes through the following steps:

1. Set $i = 0$ and generate $N$ samples $\boldsymbol{x}_0^{0,k},\ k = 0, \ldots, N$ at CL0 by standard Monte Carlo starting from the available estimate of the object state vector at the epoch $t_0$.
2. Propagate each sample for the selected time window. If $i = 0$, the time frame coincides with an arbitrarily selected investigation window (e.g. 100 years), and a list of possible

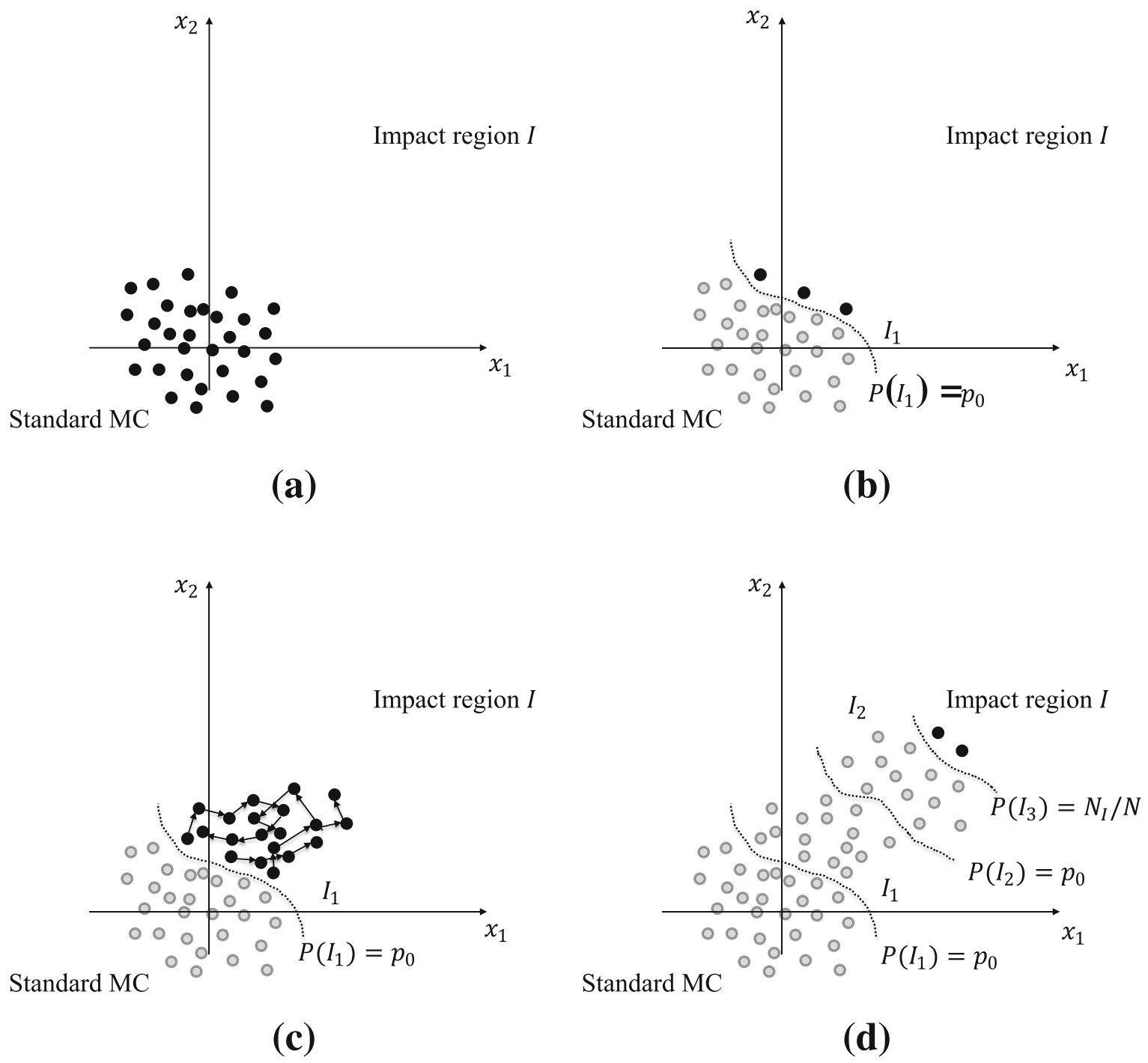

**Fig. 2** Subset simulation process: **a** initialisation by standard MC; **b** CL1 identification; **c** samples generation by means of MCMC; **d** new iterations and impact region identification

target events is obtained. Then, select one of these events $t_e$, define $\Delta t_e$ and compute the distance $d_{\Delta t_e}$. If $i > 0$, propagate each sample to $t_{e,2}$ and compute $d_{\Delta t_e}$.

3. Sort the $N$ samples in descending order according to their value of $d_{\Delta t_e}$.
4. Identify an intermediate threshold value $D_{i+1}$ as the $(1 - p_0)N$-th element of the list of $d_{\Delta t_e}$. Define the $(i + 1)$-th conditional level as $I_{i+1} = \{d_{\Delta t_e} < D_{i+1}\}$. Considering how the threshold is defined, the associated conditional probability $P(I_{i+1}|I_i) = p_0$.
5. If $D_{i+1} < D_{lim}$, where $D_{lim}$ is the target threshold distance, go the last step, otherwise select the last $p_0 N$ samples of the list $x_0^{i,j}, \; j = 0, \ldots, p_0 N$. By definition, these samples belong to the $(i + 1)$-th conditional level.
6. Using MCMC, generate $(1 - p_0)N$ additional conditional samples starting from the previously selected seeds belonging to $I_{i+1}$. A sample is set to belong to $I_{i+1}$ according to the following performance function:

$$g_X^{i+1}(\boldsymbol{x}_0) = d_{\Delta t_e}(\boldsymbol{x}_0) - D_{i+1} \begin{cases} > 0 \;\rightarrow\; \boldsymbol{x}_0 \text{ is out of the } (i+1)\text{-th CL} \\ = 0 \;\rightarrow\; \boldsymbol{x}_0 \text{ is at the limit of the } (i+1)\text{-th CL} \\ < 0 \;\rightarrow\; \boldsymbol{x}_0 \text{ belongs to the } (i+1)\text{-th CL} \end{cases}. \quad (18)$$

7. Set $i = i + 1$ and return to step 2
8. Stop the algorithm.

The total number of generated samples is

$$N_T = N + (n - 1)(1 - p_0)N, \quad (19)$$

where $n$ is the overall number of conditional levels required to reach the impact region. Since the conditional probability is equal to $p_0$ for each level, the impact probability becomes

$$\hat{P}(I) = P(I_n) = P(I_n|I_{n-1})p_0^{n-1} = p_0^{n-1}N_I/N, \tag{20}$$

where $N_I$ is the number of samples belonging to the last conditional level whose planetocentric distance is lower than $D_{lim}$.

Zuev et al. (2012) suggest a Bayesian post-processor for SS ($SS^+$) to refine the computed impact probability and determine higher moments. If we define

$$n_l = \begin{cases} p_0 N & \text{if } l < n \\ N_I & \text{if } l = n \end{cases}, \tag{21}$$

the first moment of the distribution of the impact probability becomes

$$\mathrm{E}_{SS^+}\{P\} = \prod_{l=1}^{n} \frac{n_l + 1}{N + 2}, \tag{22}$$

whereas the second moment is expressed by

$$\mathrm{E}_{SS^+}\{P^2\} = \prod_{l=1}^{n} \frac{(n_l + 1)(n_l + 2)}{(N + 2)(N + 3)}. \tag{23}$$

Therefore, the variance of the estimator can be computed as

$$\hat{\sigma}^2(P) = \mathrm{E}\{P^2\} - (\mathrm{E}\{P\})^2. \tag{24}$$

Equations 22 and 24 are the references for the analyses presented in this paper.

While the completeness in the identification of the potential impacts is influenced by the same factors mentioned in the section dedicated to the LS method, the accuracy in the characterisation of the impact probability of each identified event depends on the selection of the available degrees of freedom, i.e. the fixed conditional probability $p_0$, the number of samples per conditional level $N$ and the proposal auxiliary distribution for the generation of the samples for each MCMC phase.

The number of samples per conditional level $N$ and the conditional probability level are two key parameters and are strongly interconnected. Basically, when investigating a target event, after selecting a proper proposal distribution, suitable values for $N$ and $p_0$ need to be set and the SS process is started. As the algorithm proceeds, by knowing the imposed value of conditional probability $p_0$, the value for the currently estimated impact probability can be retrieved as $P(I) \sim p_0^n$, where $n$ is the current number of conditional levels. According to the minimum impact probability of interest $p_{min}$ (e.g. 1e−8), one can decide to stop the algorithm when $p_0^n < p_{min}$. At this point, two different outcomes are possible: if the threshold $p_{min}$ is reached, then one can say that the impact probability for the selected event is lower than the minimum probability of interest, and so a threat is ruled out. Otherwise, if the algorithm stops earlier, a positive outcome is obtained and an estimate for the impact probability is available.

As one can imagine, the reliability of the achieved results strongly depends on the entity of the selected parameter $N$. Imagine for example to target an event that is characterised by a very low probability, and to use an extremely low value of $N$, e.g. 10. It is reasonable to imagine that, unless selecting a value of $p_0$ very close to 1, thus employing a very large number of conditional levels, it would be very difficult to obtain a good result from the method. The number of samples per conditional level, therefore, directly affects how well an intermediate event region is investigated and, since the value of $p_0$ is fixed, it governs both the number of required conditional levels and the accuracy of the obtained impact probability estimate. As a result, the value of $N$ shall be compatible with the selected conditional probability $p_0$: the lower the probability, the larger the value of $N$. The most robust approach would be an iterative application of the SS, i.e. after selecting the value of $p_0$, start with a relatively low value of $N$, run the process, and if it does not give a result (i.e. if it reaches the lower probability limit without stopping), increase the value of $N$ and repeat, until the difference between two consecutive iterations is lower than an imposed threshold or no result is obtained with a preliminarily selected maximum value for $N$. This approach, however, would unavoidably weigh down the method, and it is never used in literature. On the contrary, all references generally identify an optimal value for the conditional probability $p_0$, and then suggest to tune the value of $N$ according to the complexity of the problem. In particular, since the optimal value for $p_0$ does not depend on the analysed case (see Sect. 5.2), one may decide to always select a quite large value of $N$ (e.g. 1000). This may limit the computational performance of the SS when the impact probability is large, but it prevents at least the method from giving inaccurate results. In the analyses presented in this paper, we followed this approach, by imposing a predetermined value of $p_0$ equal to 0.2, and using $N = 1000$ for all the presented test cases. A more detailed analysis on the role of these two parameters on the accuracy and efficiency of the method is offered in Sect. 5.

While $p_0$ and $N$ govern the existence of a positive outcome for the SS method, the role of the proposal distribution is equally crucial, since it rules the distribution of the candidate samples and the transition of the Markov chain from one state to another, thus affecting the variance of the estimate of the impact probability (Au and Wang 2014). Different approaches can be followed for the selection of the proposal distribution. According to (Au and Beck 2001), the efficiency of the method is insensitive to the type of the proposal pdf, hence those which can be easily operated may be used. On the other hand, the spread of the auxiliary distribution affects the size of the region covered by the Markov chain samples, thus controlling the efficiency of the method. Large spreads may reduce the acceptance rate in the Metropolis–Hastings algorithm, thus increasing the number of repeated MCMC samples. Small spreads may instead lead to high acceptance rate, but still produce very correlated samples, which negatively affects the variance. Some references suggest to maintain the same size of the original distribution, others propose an adaptive scaling based on sample statistics computed on the last conditional level (Au and Wang 2014) and acceptance rates (Zuev et al. 2012; Papaioannou et al. 2015). Essentially, two different approaches can be followed. One may decide to adopt as proposal distribution the original one or select an easy proposal distribution (e.g. Gaussian or uniform). In both cases, once defined the shape, the spread can be set either maintaining the original one or properly scaling it. In the analyses presented in this paper, we used a proposal distribution centred at the current sample, with the same size and shape of the original distribution.

## 4 Numerical simulations

This section illustrates the performance of the LS and SS methods for impact probability computation and compares the results with that of standard MC. The analyses won't deal with the process of events identification previously described for both methods, but will directly show the results achievable with well-known test cases. Three different test cases are analysed: NEAs 2010 $RF_{12}$, 2017 $RH_{16}$ and 99942 Apophis. These objects were selected because of their decreasing values of estimated impact probability as reported in the NEO risk list by the European Space Agency (ESA). Asteroid 2010 $RF_{12}$ is a small NEA that currently has the highest estimated probability of hitting the Earth ($\approx 6.50 \cdot 10^{-2}$) in 2095. Asteroid 2017 $RH_{16}$ is the second object considered in our analysis. According to the initial conditions dating back to 24 September 2017, this object would be characterised by an estimated impact probability in 2026 of $\approx 1.45 \cdot 10^{-3}$. Though recently ruled out with new observations, this second case offers an interesting scenario as it is characterised by a much shorter propagation window and a lower impact probability than the case of asteroid 2010 $RF_{12}$. The last case is asteroid 99942 Apophis. For this case, in order to test the two methods on the worst scenario, the initial conditions are sampled from the estimate obtained by discarding observations performed after 2009. In such a scenario, after a very close encounter in 2029, the asteroid might experience a resonant return with the Earth in 2036, with an estimated impact probability of $3 \cdot 10^{-5}$ (Sciarra 2020).

For all the considered test cases, the initial conditions are obtained in terms of equinoctial parameters (Broucke and Cefola 1972) and then converted into Cartesian coordinates for the propagation. All the required data were retrieved in July 2018 from the Near-Earth Object Dynamic Site (https://newton.spacedys.com/neodys/) and the NEO Risk List from ESA (http://neo.ssa.esa.int/risk-page), and are reported in Tables 1, 2 and 3. The data for asteroid Apophis were instead provided by prof. G. Tommei from the University of Pisa (personal communication, 2018). For the first two cases, the potential inaccuracies deriving from the assumption of linearity for the uncertainty region generated by the large time gap between the data arc and the osculating epoch are not considered.

The propagations are carried out in Cartesian coordinates with respect to the J2000 reference frame centred at the Solar System Barycentre (SSB), with the inclusion of the gravitational contributions of the Sun, all the major planets, and the Earth's moon, including relativistic effects modelled as in Armellin et al. (2010). All the physical constants (gravitational parameters, planetary radii, etc.) and the ephemerides are obtained from the JPL Horizons database via the SPICE toolkit (https://naif.jpl.nasa.gov/naif/). All propagations are performed in normalised units (with reference length and time equal to 1 au and 1 solar year, respectively) using the adaptive Dormand–Prince Runge–Kutta scheme of $8^{\text{th}}$ order (RK78, Prince and Dormand (1981)), with absolute and relative tolerances both set to $10^{-12}$.

The comparison between the standard MC and the two proposed approaches is performed by analysing six parameters:

- the number of random initial conditions $N_{IC}$ (equal to the number of lines in the LS, and to the number of samples per conditional level in the SS);
- the total number of orbital propagations $N_P$ (larger than $N_{IC}$ for both LS, due to the implemented iterative procedure, and SS, due to the required conditional levels);
- the impact probability estimate $\hat{P}(I)$;
- the sample standard deviation $\hat{\sigma}$ of $\hat{P}(I)$;
- the coefficient of variation $\delta$ of the probability estimate, defined as $\hat{\sigma}/\hat{P}(I)$;
- the figure of merit (FoM) of the method, defined as $1/(\hat{\sigma}^2 \cdot N_P)$ following the notation defined in Zio and Pedroni (2009).

In the case of LS, $N_P$ also includes the extra orbital propagations that are used in the preliminary phases (optimisation and MCMC). This parameter is selected as a measure of the computational burden of the methods. The standard deviation and the coefficient of variation are instead used as indicators of the accuracy of the result, with lower values corresponding to lower variability. The FoM involves both the variance and the required number of propagations, and it is a measure of the computational efficiency and impact probability variability: the higher the value, the higher the efficiency of the method(Zio and Pedroni 2009; Morselli et al. 2015).

**Table 1** Nominal equinoctial parameters (second row) and related covariance matrix (last six rows) for asteroid 2010 $RF_{12}$ at 6656 MJD (TDT) (23 March 2018)

| $a$ (AU) | $P_1(-)$ | $P_2(-)$ | $Q_1(-)$ | $Q_2(-)$ | $l$ (∘) |
|---|---|---|---|---|---|
| 1.060446078146929 | 0.178466224199297 | 0.060022015835624 | 0.002143975938021 | −0.007400721981213 | 331.3598212963466 |
| $1.8261{\cdot}10^{-11}$ | $-4.5772{\cdot}10^{-11}$ | $2.7901{\cdot}10^{-11}$ | $-4.344{\cdot}10^{-12}$ | $1.6851{\cdot}10^{-11}$ | $-7.2039142{\cdot}10^{-08}$ |
| $-4.5772{\cdot}10^{-11}$ | $1.18764{\cdot}10^{-10}$ | $-7.0652{\cdot}10^{-11}$ | $1.1136{\cdot}10^{-11}$ | $-4.3212{\cdot}10^{-11}$ | $1.81110710{\cdot}10^{-07}$ |
| $2.7901{\cdot}10^{-11}$ | $-7.0652{\cdot}10^{-11}$ | $4.2758{\cdot}10^{-11}$ | $-6.682{\cdot}10^{-12}$ | $2.5920{\cdot}10^{-11}$ | $-1.10166354{\cdot}10^{-07}$ |
| $-4.344{\cdot}10^{-12}$ | $1.1136{\cdot}10^{-11}$ | $-6.682{\cdot}10^{-12}$ | $1.049{\cdot}10^{-12}$ | $-4.071{\cdot}10^{-12}$ | $1.7171456{\cdot}10^{-08}$ |
| $1.6851{\cdot}10^{-11}$ | $-4.3212{\cdot}10^{-11}$ | $2.5920{\cdot}10^{-11}$ | $-4.071{\cdot}10^{-12}$ | $1.5793{\cdot}10^{-11}$ | $-6.6608995{\cdot}10^{-08}$ |
| $-7.2039142{\cdot}10^{-08}$ | $1.81110710{\cdot}10^{-07}$ | $-1.10166354{\cdot}10^{-07}$ | $1.7171456{\cdot}10^{-08}$ | $-6.6608995{\cdot}10^{-08}$ | $2.84264603280{\cdot}10^{-04}$ |

**Table 2** Nominal equinoctial parameters (second row) and related covariance matrix (last six rows) for asteroid 2017 $RH_{16}$ at 6475 MJD (TDT) (24 September 2017)

| $a$ (AU) | $P_1(-)$ | $P_2(-)$ | $Q_1(-)$ | $Q_2(-)$ | $l$ (∘) |
|---|---|---|---|---|---|
| 0.875165560919210 | −0.186708542625558 | −0.401438413356919 | −0.001983912479439 | 0.005025249763243 | 319.9653304201169 |
| $9.7909837 \cdot 10^{-08}$ | $1.2615584 \cdot 10^{-08}$ | $3.6644068 \cdot 10^{-08}$ | $-1.106735 \cdot 10^{-09}$ | $-2.22775 \cdot 10^{-10}$ | $-2.274493142 \cdot 10^{-06}$ |
| $1.2615584 \cdot 10^{-08}$ | $1.874471 \cdot 10^{-09}$ | $5.232931 \cdot 10^{-09}$ | $-1.39989 \cdot 10^{-10}$ | $-3.7340 \cdot 10^{-11}$ | $-2.19893323 \cdot 10^{-07}$ |
| $3.6644 \cdot 10^{-08}$ | $5.232931 \cdot 10^{-09}$ | $1.4767424 \cdot 10^{-08}$ | $-4.08824 \cdot 10^{-10}$ | $-1.01142 \cdot 10^{-10}$ | $-7.00731267 \cdot 10^{-07}$ |
| $-1.106735 \cdot 10^{-09}$ | $-1.39989 \cdot 10^{-10}$ | $-4.08824 \cdot 10^{-10}$ | $1.2541 \cdot 10^{-11}$ | $2.408 \cdot 10^{-12}$ | $2.6479634 \cdot 10^{-08}$ |
| $-2.22775 \cdot 10^{-10}$ | $-3.7340 \cdot 10^{-11}$ | $-1.01142 \cdot 10^{-10}$ | $2.408 \cdot 10^{-12}$ | $9.21 \cdot 10^{-13}$ | $2.634567 \cdot 10^{-09}$ |
| $-2.274493142 \cdot 10^{-06}$ | $-2.19893323 \cdot 10^{-07}$ | $-7.00731267 \cdot 10^{-07}$ | $2.6479634 \cdot 10^{-08}$ | $2.634567 \cdot 10^{-09}$ | $7.4364743783 \cdot 10^{-05}$ |

**Table 3** Nominal equinoctial parameters (second row) and related covariance matrix (last six rows) for asteroid 99942 Apophis at 1820 MJD (TDT) (25 December 2004)

| $a$ (AU) | $P_1(-)$ | $P_2(-)$ | $Q_1(-)$ | $Q_2(-)$ | $l$ (∘) |
|---|---|---|---|---|---|
| 0.922194734054135 | − 0.093240000935568 | 0.166968931027762 | − 0.012074399344515 | − 0.026456592668282 | 68.0924709576347 |
| $3.622806{\cdot}10^{-14}$ | $-1.221227{\cdot}10^{-14}$ | $1.0569229{\cdot}10^{-13}$ | $2.58102{\cdot}10^{-15}$ | $-2.107900{\cdot}10^{-14}$ | $-1.731178991{\cdot}10^{-11}$ |
| $-1.221227{\cdot}10^{-14}$ | $2.1708118{\cdot}10^{-13}$ | $-4.6053117{\cdot}10^{-13}$ | $1.205162{\cdot}10^{-14}$ | $8.125138{\cdot}10^{-14}$ | $7.595377366{\cdot}10^{-11}$ |
| $1.0569229{\cdot}10^{-13}$ | $-4.6053117{\cdot}10^{-13}$ | $1.16445867{\cdot}10^{-12}$ | $-1.939228{\cdot}10^{-14}$ | $-2.0862937{\cdot}10^{-13}$ | $-1.9127667143{\cdot}10^{-10}$ |
| $2.58102{\cdot}10^{-15}$ | $1.205162{\cdot}10^{-14}$ | $-1.939228{\cdot}10^{-14}$ | $6.40085{\cdot}10^{-15}$ | $1.85544{\cdot}10^{-15}$ | $3.05473374{\cdot}10^{-12}$ |
| $-2.107900{\cdot}10^{-14}$ | $8.125138{\cdot}10^{-14}$ | $-2.0862937{\cdot}10^{-13}$ | $1.85544{\cdot}10^{-15}$ | $3.860434{\cdot}10^{-14}$ | $3.445555421{\cdot}10^{-11}$ |
| $-1.731178991{\cdot}10^{-11}$ | $7.595377366{\cdot}10^{-11}$ | $-1.9127667143{\cdot}10^{-10}$ | $-1.9127667143{\cdot}10^{-10}$ | $3.05473374{\cdot}10^{-12}$ | $3.445555421{\cdot}10^{-11}$ |

**Fig. 3** Samples dispersion in the initial uncertainty space ($\delta a$,$\delta l$) for the case of asteroid 2010 $RF_{12}$: **a** initial conditions leading to impact obtained via standard MC; **b** boundaries of the subdomain identified via LS; **c** samples per conditional level obtained with SS

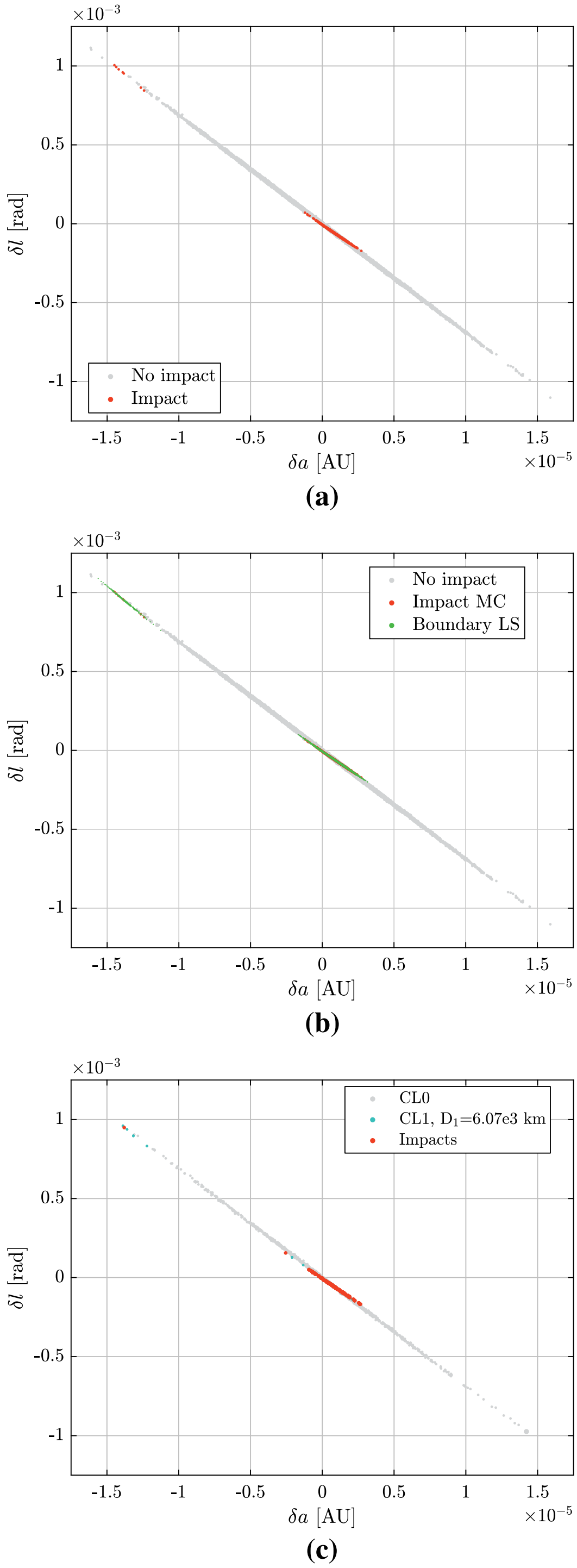

**Table 4** Performance comparison between standard MC, LS and SS when applied to the case of asteroid 2010 $RF_{12}$

| | $N_{IC}$ | $N_P$ | $\hat{P}(I)$ | $\hat{\sigma}(P(I))$ | $\delta$ | FoM |
|---|---|---|---|---|---|---|
| MC | 10000 | 10000 | $6.51{\cdot}10^{-2}$ | $2.47{\cdot}10^{-3}$ | $3.79{\cdot}10^{-2}$ | 16.39 |
| LS (ref) | 1000 | 8174 | $6.57{\cdot}10^{-2}$ | $3.30{\cdot}10^{-3}$ | $5.02{\cdot}10^{-2}$ | 11.23 |
| LS ($\sigma^{MC}$) | 1800 | 14574 | $6.68{\cdot}10^{-2}$ | $2.48{\cdot}10^{-3}$ | $3.71{\cdot}10^{-2}$ | 11.16 |
| LS ($N_P^{MC}$) | 1250 | 10174 | $6.58{\cdot}10^{-2}$ | $2.98{\cdot}10^{-3}$ | $4.53{\cdot}10^{-2}$ | 11.07 |
| SS (ref) | 1000 | 1900 | $6.53{\cdot}10^{-2}$ | $6.36{\cdot}10^{-3}$ | $9.72{\cdot}10^{-2}$ | 13.01 |
| SS ($\sigma^{MC}$) | 6500 | 12350 | $6.49{\cdot}10^{-2}$ | $2.48{\cdot}10^{-3}$ | $3.82{\cdot}10^{-2}$ | 13.17 |
| SS ($N_P^{MC}$) | 5000 | 9500 | $6.49{\cdot}10^{-2}$ | $2.83{\cdot}10^{-3}$ | $4.36{\cdot}10^{-2}$ | 13.14 |

Table 1 shows the initial conditions in terms of nominal equinoctial parameters and related uncertainties for asteroid 2010 $RF_{12}$ on 23 March 2018.

Figure 3a shows the result of a standard MC sampling of the initial uncertainty set for the asteroid in terms of deviations of semi-major axis and equinoctial longitude from the nominal initial conditions, with red dots representing initial conditions leading to an impact at the investigated epoch. Table 4 summarises the numerical results of the simulation: the resulting impact probability as obtained by propagating 10000 samples is $6.51{\cdot}10^{-2}$, whereas the estimated Poisson statistics uncertainty is $2.47{\cdot}10^{-3}$.

Figure 3b shows the results obtained with the LS method. The grey dots are samples from the initial distribution that do not lead to impact, green dots are the initial conditions lying on the boundaries of the impact region as identified by the LS algorithm, red dots identify initial conditions leading to an impact. Using 1000 initial samples with 8 propagations for the iterative process along each sampling line, the resulting probability is $6.57{\cdot}10^{-2}$, whereas the estimated uncertainty is $3.30{\cdot}10^{-3}$, as shown in Table 4. Figure 3c, instead, shows the evolution of the conditional samples obtained with SS. The method is applied with 1000 samples per conditional level, a fixed conditional probability equal to 0.2, and an auxiliary distribution centred in the current sample and with the same magnitude of the original one. In the plot, grey dots represent samples drawn at CL0 by standard MC, whereas green dots belong to CL1. Among these, impacting samples are represented in red. As can be seen, only two conditional levels are generated to obtain an impact probability of $6.53{\cdot}10^{-2}$.

Table 4 shows a comparison between the three methods. The comparison is done with respect to the performance indexes described at the beginning of this section. Along with the results illustrated in Fig. 3 (here referred to as "ref"), two additional simulations are presented for the LS and SS methods: the results obtained using a number of samples granting the same accuracy level of standard MC ($\sigma^{MC}$ approach), and the results obtained by performing the same number of propagations of the standard MC ($N_P^{MC}$ approach). Let us first compare the results of standard MC with the reference results obtained with LS and SS. As can be seen, though both LS and SS provide an impact probability estimate that is very close to the MC result with a lower number of propagations, their overall performance is lower, as confirmed by the larger value for $\delta$ and the lower figure of merit FoM. This trend becomes more clear if the results obtained with the $\sigma^{MC}$ and $N_P^{MC}$ approaches are analysed. As can be seen, for the analysed test case, neither the LS nor the SS offer a real advantage with respect to the standard

**Table 5** Performance comparison between standard MC, LS and SS when applied to the case of asteroid 2017 $RH_{16}$

| | $N_{IC}$ | $N_P$ | $\hat{P}(I)$ | $\hat{\sigma}(P(I))$ | $\delta$ | FoM |
|---|---|---|---|---|---|---|
| MC | 50000 | 50000 | $1.42{\cdot}10^{-3}$ | $1.68{\cdot}10^{-4}$ | $1.18{\cdot}10^{-1}$ | 708.62 |
| LS (ref) | 1000 | 8245 | $1.56{\cdot}10^{-3}$ | $7.19{\cdot}10^{-5}$ | $4.61{\cdot}10^{-2}$ | $2.35{\cdot}10^{4}$ |
| LS ($\sigma^{MC}$) | 164 | 1557 | $1.43{\cdot}10^{-3}$ | $1.70{\cdot}10^{-4}$ | $1.19{\cdot}10^{-1}$ | $2.22{\cdot}10^{4}$ |
| LS ($N_P^{MC}$) | 6250 | 50245 | $1.42{\cdot}10^{-3}$ | $2.85{\cdot}10^{-5}$ | $2.01{\cdot}10^{-2}$ | $2.45{\cdot}10^{4}$ |
| SS (ref) | 1000 | 2800 | $1.45{\cdot}10^{-3}$ | $2.24{\cdot}10^{-4}$ | $1.54{\cdot}10^{-1}$ | 7117.80 |
| SS ($\sigma^{MC}$) | 2000 | 5600 | $1.43{\cdot}10^{-3}$ | $1.57{\cdot}10^{-4}$ | $1.10{\cdot}10^{-1}$ | 7244.57 |
| SS ($N_P^{MC}$) | 18000 | 50400 | $1.42{\cdot}10^{-3}$ | $5.20{\cdot}10^{-5}$ | $3.66{\cdot}10^{-2}$ | 7337.75 |

MC: the latter guarantees the same accuracy level with a lower number of propagations and higher accuracy with the same number of propagations. This result is strictly related to the relatively high value of the estimated impact probability, as shown in Table 4. Essentially, the reduction in the number of propagations granted by both advanced MC method does not compensate for the increased uncertainty their estimates are affected by, thus leading to the results shown in Table 4.

The performance of LS and SS becomes more interesting as the probability of the event decreases. Table 2 reports the initial conditions in terms of nominal equinoctial parameters and associated uncertainties for the second test case, asteroid 2017 $RH_{16}$, whereas Fig. 4 shows the results of the numerical simulations obtained with the three methods.

The MC result shown in Fig. 4a is obtained by propagating 50000 samples drawn from the initial distribution, and the estimated impact probability is equal to $1.42{\cdot}10^{-3}$. Figure 4b, instead, shows the result of the LS method. Similarly to the previous case, 1000 lines are used to characterise the impact regions, and the resulting impact probability is equal to $1.56{\cdot}10^{-3}$. Finally, Fig. 4c shows the outcome of the SS method. The same setting parameters of the previous test case are used here, but a different random walk in the space of the initial conditions is obtained. As can be seen, one extra conditional level is required, leading to an estimated probability of $1.45{\cdot}10^{-3}$. An overview of the obtained performance is shown in Table 5. The estimated probability is now one order of magnitude lower than the one associated with asteroid 2010 $RF_{12}$. In this case, as shown in Table 5, both LS and SS outperform standard MC in terms of both achieved accuracy and computational cost. More in detail, the same accuracy level of standard MC can be obtained by both LS and SS with a number of propagations that is one order of magnitude lower. Alternatively, LS and SS achieve a higher accuracy level with the same number of samples adopted for the standard MC. These trends become quite evident from the analysis of the $\delta$ and FoM parameters. In particular, with the $\sigma^{MC}$ approach, both LS and SS employ a lower number of samples, and so larger values of FoM are obtained. On the other hand, if the same number of propagations is used, both methods guarantee a reduction in the probability estimate uncertainty of at least a factor 2, thus resulting in lower values of $\delta$ and much larger values for the parameter FoM.

The positive trends previously described become more evident when very rare events are investigated. Table 3 shows the initial conditions in terms of nominal equinoctial parameters and related uncertainties for the third test case, asteroid 99942 Apophis. This test case offers the most challenging scenario due to the presence of the deep close encounter in 2029, which triggers the resonant return in 2036.

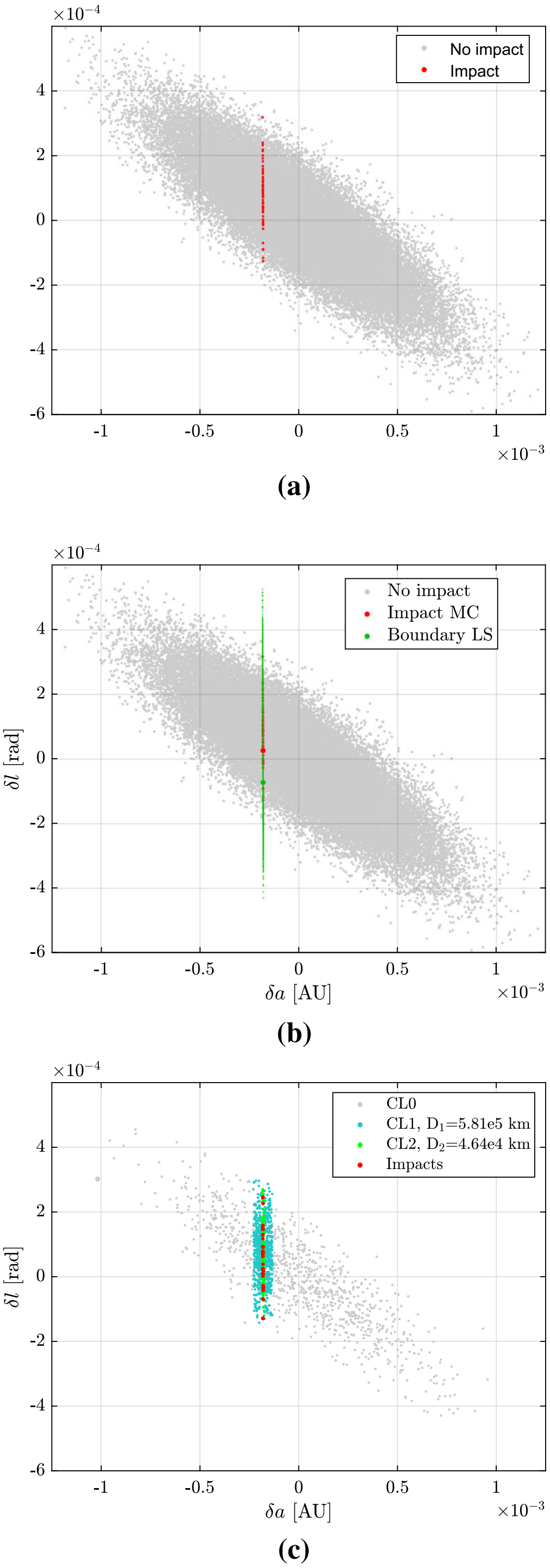

**Fig. 4** Samples dispersion in the initial uncertainty space ($\delta a$,$\delta l$) for the case of asteroid 2017 $RH_{16}$: **a** initial conditions leading to impact obtained via standard MC; **b** boundaries of the subdomain identified via LS; **c** samples per conditional level obtained with SS

Figure 5a shows the distribution of the initial conditions and the impacting samples as obtained with the standard MC method. One million samples are used, resulting into an estimated impact probability of $3.00 \cdot 10^{-5}$. The significantly lower probability unavoidably affects also the two proposed methods. This is particularly evident for the SS method (Fig. 5c), which requires a larger number of conditional levels. In the end, by setting the same parameters of the previous two examples, their estimated impact probabilities are equal to $3.12 \cdot 10^{-5}$ and $3.36 \cdot 10^{-5}$, respectively. Table 6 reports the results of the numerical simulations obtained with the three methods. The performance comparison confirms the benefits provided by both LS and SS over the standard MC. As can be seen, reliable results with the same accuracy level of standard MC can be obtained by using a number of samples three orders of magnitude lower. On the other hand, a reduction of at least one order of magnitude in the estimate uncertainty can be obtained by propagating the same number of samples of a direct MC approach. This final example allows us also to assess the relative performance of the two proposed approaches: while SS is very efficient in providing reliable impact probability results for relatively low number of samples, its performance does not significantly change for increasing numbers of samples per conditional level. This trend is valid also for the LS method, which however requires almost always a larger number of propagations for converging but provides impact probability estimates characterised by much smaller levels of uncertainty.

Figure 6 shows a summary of the obtained results. The two plots show the results obtained in terms of normalised coefficient of variation (Fig. 6a) and FoM (Fig. 6b) as a function of the estimated impact probability for the three analysed test cases. Normalisation is performed with respect to the results of direct MC. Triangles and squares are used for the LS and SS methods, respectively, whereas different colours refer to the different selected approaches: black for the reference, grey for the $\sigma^{MC}$ approach and light grey for the $N_P^{MC}$ approach. Let us first analyse Fig. 6a, and let us consider the results obtained with the reference simulation (black). As can be seen, while for the first test case both LS and SS estimates show larger uncertainties with respect to direct MC, as the estimated impact probability decreases, the associated uncertainty provided by the LS and the SS becomes smaller than the one provided by standard MC, with an increasingly larger gap between the two methods in favour of the LS method. This trend is enhanced if the same number of samples used for the MC simulations is employed. Similar considerations can be made if the trend of the normalised FoM is analysed. In particular, the difference between the two methods becomes clearer: for decreasing values of impact probability, the LS method provides values of the FoM that are increasingly better than the ones obtained with the SS. That is, the larger number of samples generally required by the LS method is highly compensated by the significantly lower uncertainty of the obtained estimates.

The presented analyses offer a general overview of the performance of the two proposed advanced MC method for NEA impact probability computation. The three analysed scenarios show that both LS and SS become increasingly attractive as the probability of the target failure event becomes lower, with a reduction in the number of required propagation of some orders of magnitude. On the other hand, when the impact probability is larger, standard MC still offers competitive performance. This trend, which was already highlighted by Morselli et al. (2015) for conjunctions analysis, is confirmed also for NEA impact probability computation. In real-world scenarios, the case of low or very low impact probabilities represents the most common situation, which suggests that these methods may be valuable tools for impact monitoring.

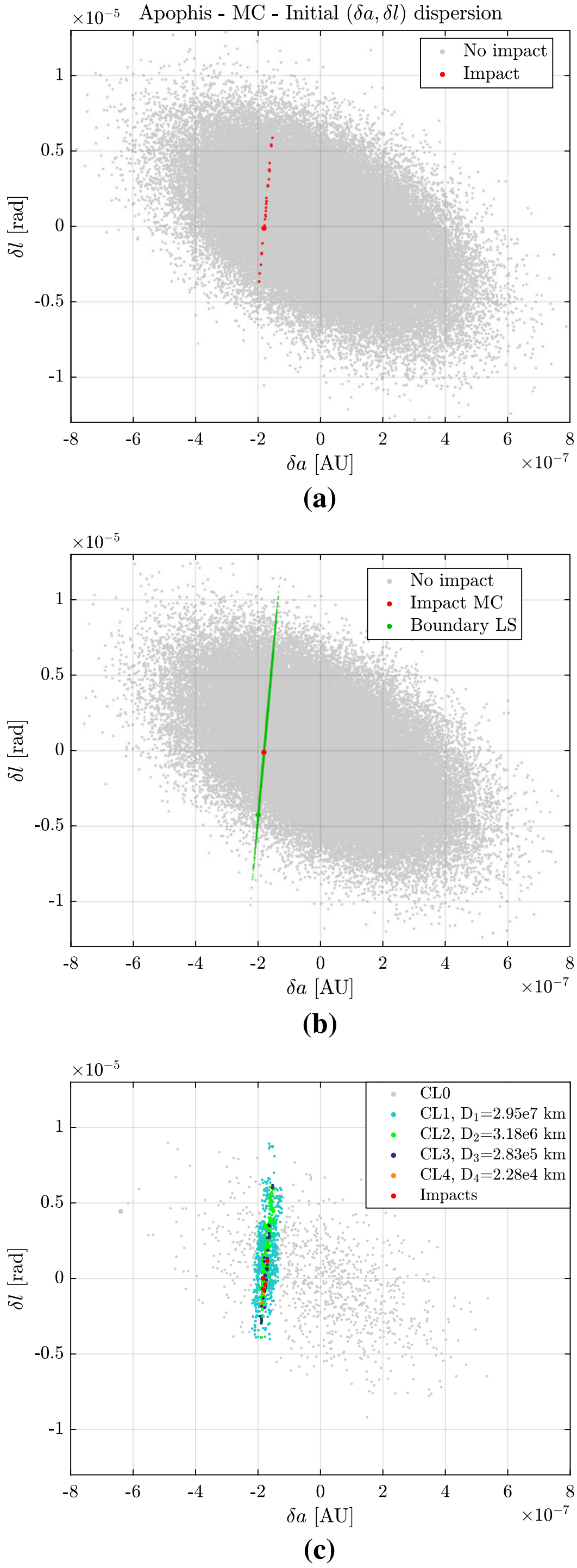

**Fig. 5** Samples dispersion in the initial uncertainty space ($\delta a$,$\delta l$) for the case of asteroid 99942 Apophis: **a** initial conditions leading to impact obtained via standard MC; **b** boundaries of the subdomain identified via LS; **c** samples per conditional level obtained with SS

**Table 6** Performance comparison between standard MC, LS and SS when applied to the case of asteroid 99942 Apophis

| | $N_{IC}$ | $N_P$ | $\hat{P}(I)$ | $\hat{\sigma}(P(I))$ | $\delta$ | FoM |
|---|---|---|---|---|---|---|
| MC | 1000000 | 1000000 | $3.00 \cdot 10^{-5}$ | $5.48 \cdot 10^{-6}$ | $1.83 \cdot 10^{-1}$ | $3.33 \cdot 10^{4}$ |
| LS (ref) | 1000 | 10351 | $3.12 \cdot 10^{-5}$ | $2.64 \cdot 10^{-7}$ | $8.46 \cdot 10^{-3}$ | $1.39 \cdot 10^{9}$ |
| LS ($\sigma^{MC}$) | 216 | 2511 | $3.19 \cdot 10^{-5}$ | $5.48 \cdot 10^{-6}$ | $1.72 \cdot 10^{-1}$ | $1.33 \cdot 10^{7}$ |
| LS ($N_P^{MC}$) | 100000 | 1000351 | $3.10 \cdot 10^{-5}$ | $2.56 \cdot 10^{-8}$ | $8.26 \cdot 10^{-4}$ | $1.53 \cdot 10^{9}$ |
| SS (ref) | 1000 | 4600 | $3.36 \cdot 10^{-5}$ | $6.56 \cdot 10^{-6}$ | $1.95 \cdot 10^{-1}$ | $5.05 \cdot 10^{6}$ |
| SS ($\sigma^{MC}$) | 1500 | 6900 | $3.27 \cdot 10^{-5}$ | $5.30 \cdot 10^{-6}$ | $1.62 \cdot 10^{-1}$ | $5.16 \cdot 10^{6}$ |
| SS ($N_P^{MC}$) | 220000 | 1010000 | $3.15 \cdot 10^{-5}$ | $5.22 \cdot 10^{-7}$ | $1.66 \cdot 10^{-2}$ | $4.72 \cdot 10^{6}$ |

## 5 Sensitivity analysis

The performance of the LS and SS is strongly affected by the selection of the available degrees of freedom, which govern both the existence and the accuracy of the solution. This section reports an analysis of the sensitivity of the LS and SS performance to the different setting parameters. For both methods, the analysis is carried out in terms of the performance criteria defined in Sect. 4.

### 5.1 Line sampling

As described in Sect. 2, the determination of the sampling direction and the identification of the boundaries of the impact region are the two main aspects of the LS method. On one hand, the sampling direction is the main parameter that affects the accuracy of LS: the method is able to achieve an accuracy level that is at least equal to the one obtained via standard MC, with the optimal case represented by a sampling direction that is orthogonal to the boundary of the impact region Zio (2013). On the other hand, the value of the probability estimate depends on the identification of the impact region, as the intersection between its boundary and the sampling lines directly affects the evaluation of the probability integrals. While there are many parameters influencing the performance of LS (the determination of the starting point, the number of elements of the MCMC, the method itself to determine the sampling direction, the numerical method used to identify the roots of the performance function while sampling along each line, its stopping conditions, etc.), these degrees of freedom all resolve into the two main aspects that are here addressed: the sampling direction itself and the determination of the boundary of the impact region. In the following paragraphs, the effect of these two aspects on the performance of the method is investigated.

#### 5.1.1 Sampling direction

The reference direction is the main parameter affecting the numerical performance of the line sampling method, since the subsequent sampling procedure for the identification of the boundaries of the impact region depends on it. As already mentioned in Sect. 2.2, this direction can be determined in different ways, which are introduced in Zio and Pedroni (2009).

For this analysis, the reference direction is first identified with the procedure in Sect. 2.2 and then slightly perturbed. The NEAs 2010 $RF_{12}$ and 99942 Apophis are used as reference

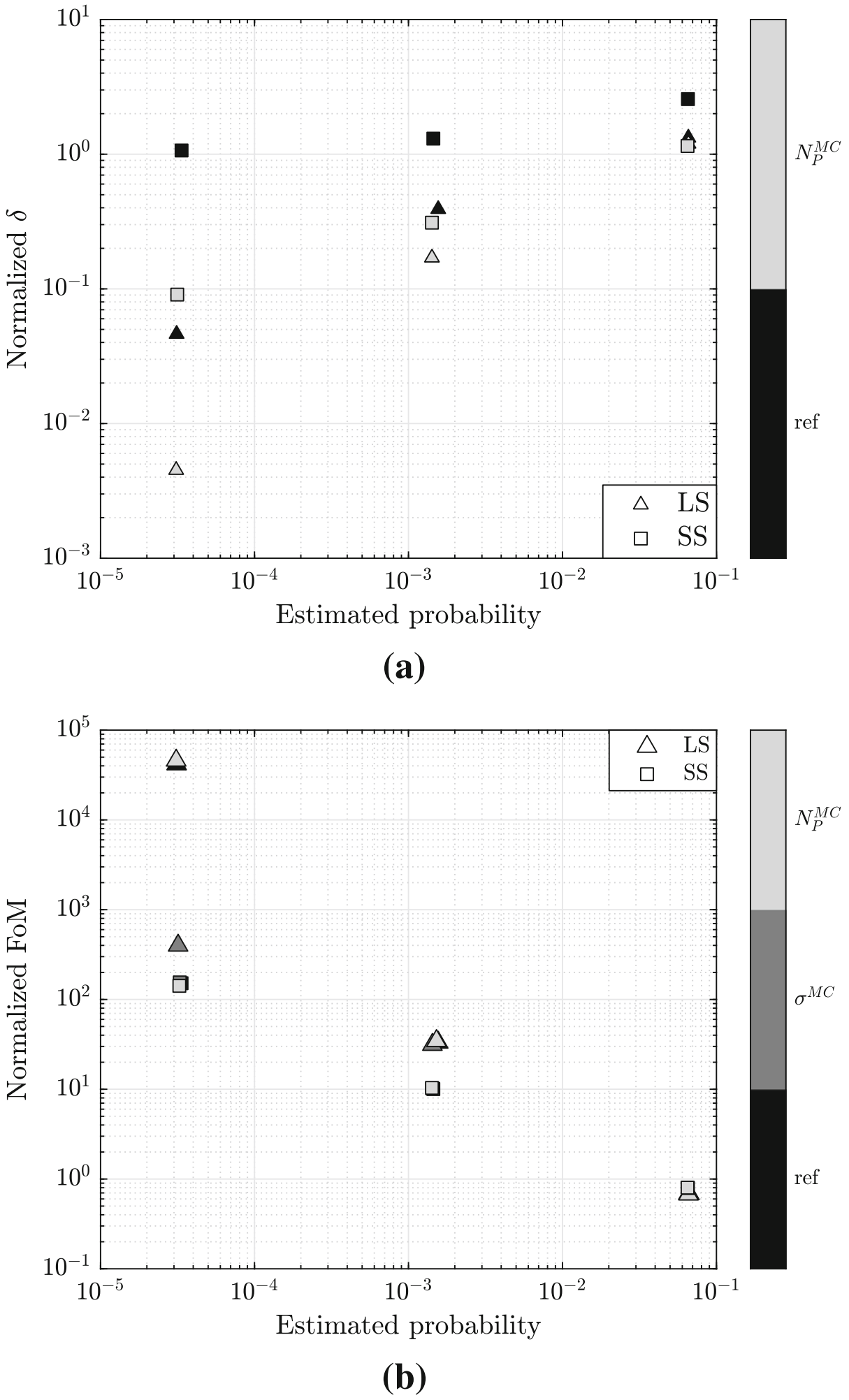

**Fig. 6** Performance parameters comparison for the three different test cases: **a** normalised coefficient of variations and **b** figure of merit as a function of the estimated impact probability. Normalisation is done with respect to the value of the direct MC. Triangles and squares are used for the LS and SS methods, respectively. Different colours are used for the different approaches (reference, $\sigma^{MC}$ and $N_P^{MC}$)

test cases, since they feature the highest and lowest impact probability, respectively, among all the cases addressed in Sect. 4. In the examples, the deviation of the $\boldsymbol{\alpha}$ direction is obtained by adding a component of arbitrary size orthogonal to the nominal direction in the ($\delta a$,$\delta l$) space (the same outcome could be obtained by using a lower number of points in the initial Markov chain, resulting in an uneven coverage of the impact region).

Figures 7 and 8 show the samples distributions obtained with the LS in the two test cases, whereas Tables 7 and 8 summarise the results. As can be seen, the LS performance drops off in the perturbed case, since the method shows larger standard deviations and lower FoMs with respect to the results presented in Sect. 4 (reported again in these tables for reader's convenience). The performance penalty is contained for 2010 $RF_{12}$, due to the high impact probability and because the sampling lines still intersect most of the impact region (see Fig. 7). For 99942 Apophis, the penalty is more relevant, as a larger number of samples is needed to converge to a solution because both the impact probability is lower and fewer sampling lines intersect the impact region.

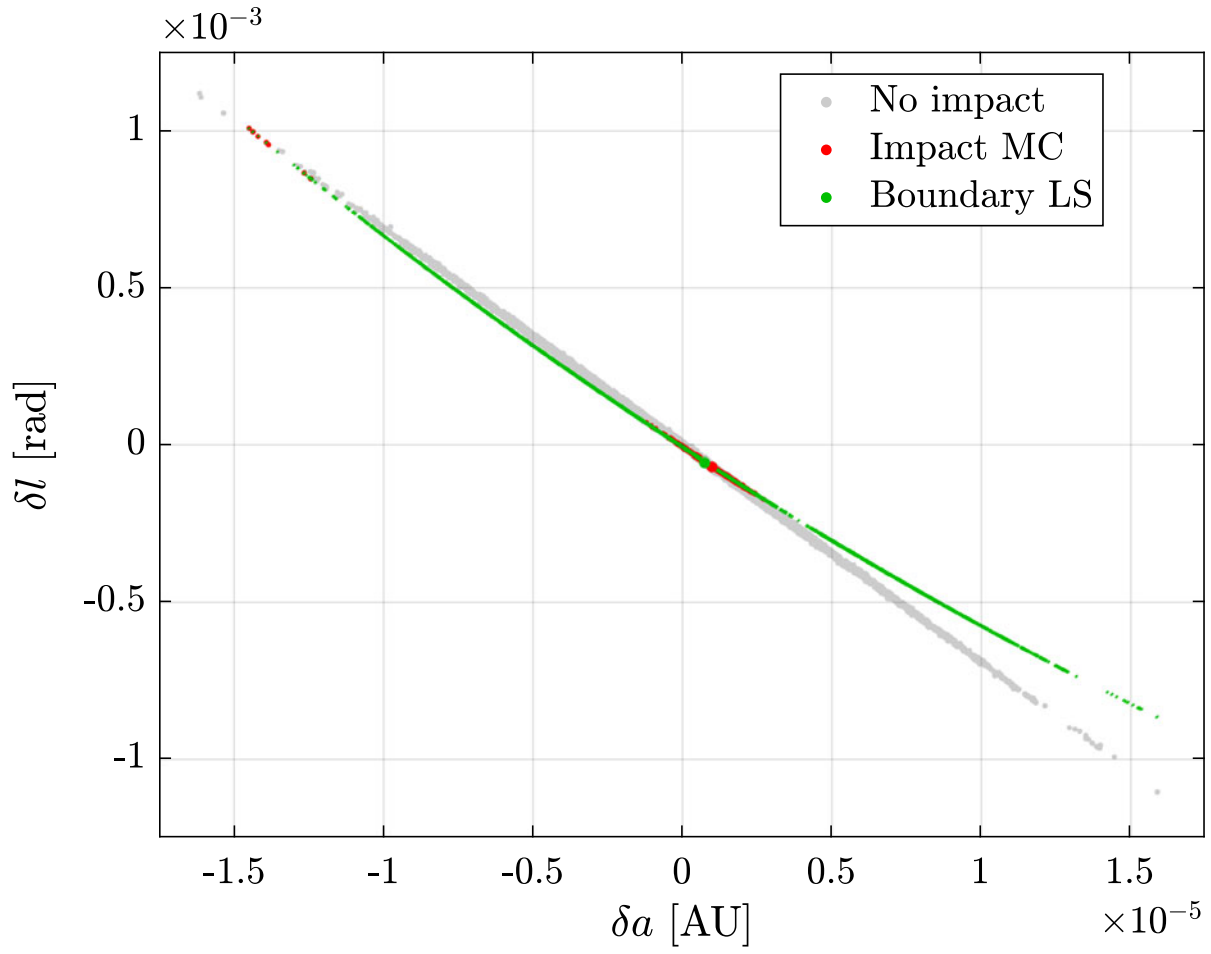

**Fig. 7** Samples dispersion in the initial uncertainty space ($\delta a$,$\delta l$) for the case of asteroid 2010 $RF_{12}$ when the LS method is applied with a perturbed sampling direction

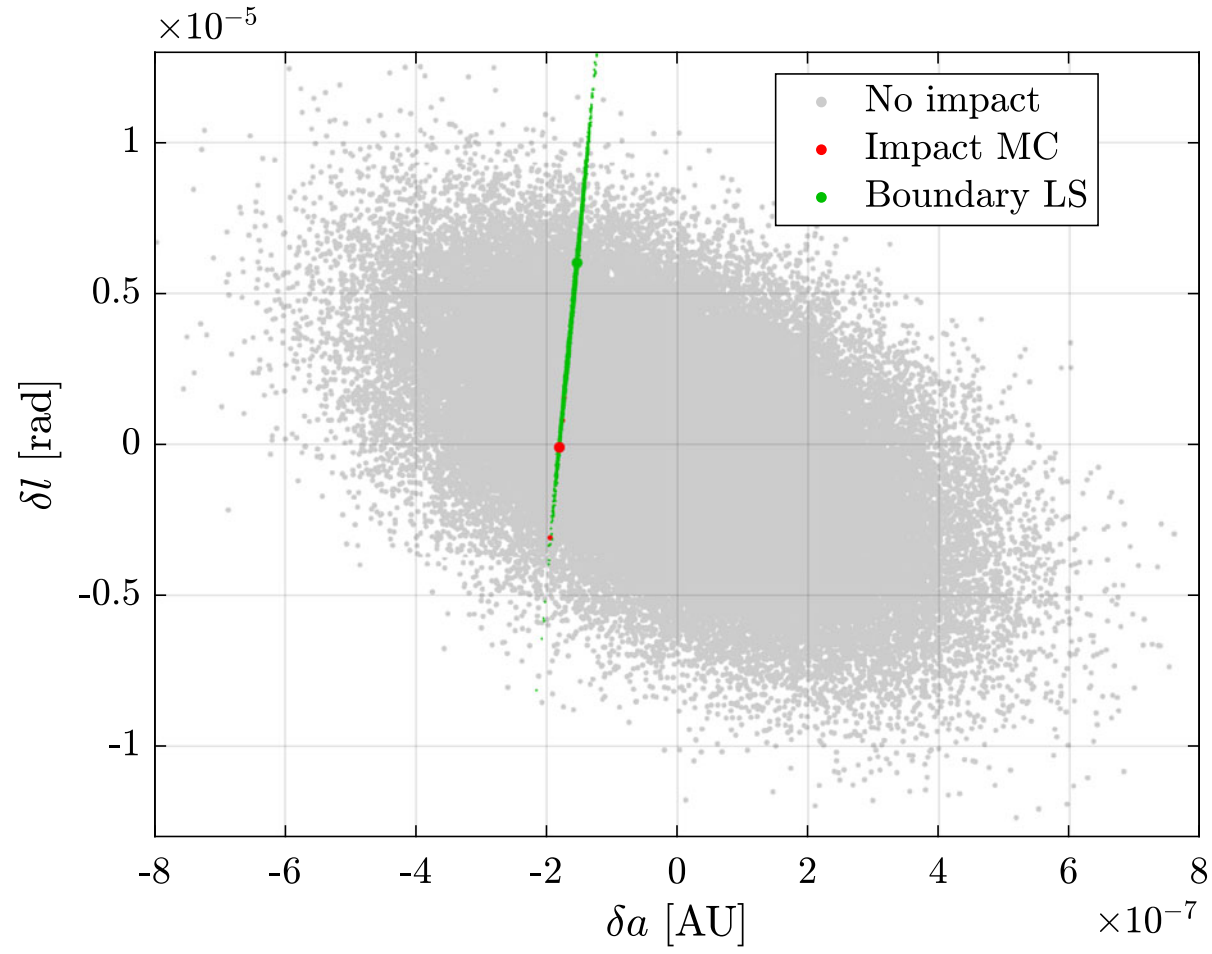

**Fig. 8** Samples dispersion in the initial uncertainty space ($\delta a$,$\delta l$) for the case of asteroid 99942 Apophis when the LS method is applied with a perturbed sampling direction

**Table 7** Application of LS with a perturbed sampling direction to the case of asteroid 2010 $RF_{12}$

| | $N_{IC}$ | $N_P$ | $\hat{P}(I)$ | $\hat{\sigma}(P(I))$ | $\delta$ | FoM |
|---|---|---|---|---|---|---|
| MC | 10000 | 10000 | $6.51{\cdot}10^{-2}$ | $2.47{\cdot}10^{-3}$ | $3.79{\cdot}10^{-2}$ | 16.43 |
| $LS_{nom}$ (ref) | 1000 | 8174 | $6.57{\cdot}10^{-2}$ | $3.30{\cdot}10^{-3}$ | $5.02{\cdot}10^{-2}$ | 11.22 |
| $LS_{nom}$ ($\sigma^{MC}$) | 1800 | 14574 | $6.68{\cdot}10^{-2}$ | $2.48{\cdot}10^{-3}$ | $3.71{\cdot}10^{-2}$ | 11.17 |
| $LS_{nom}$ ($N_P^{MC}$) | 1250 | 10174 | $6.58{\cdot}10^{-2}$ | $2.98{\cdot}10^{-3}$ | $4.53{\cdot}10^{-2}$ | 11.08 |
| $LS_{pert}$ (ref) | 1200 | 9752 | $6.68{\cdot}10^{-2}$ | $4.68{\cdot}10^{-3}$ | $7.01{\cdot}10^{-2}$ | 4.68 |
| $LS_{pert}$ ($\sigma^{MC}$) | 4332 | 34808 | $6.72{\cdot}10^{-2}$ | $2.47{\cdot}10^{-3}$ | $3.68{\cdot}10^{-2}$ | 4.70 |
| $LS_{pert}$ ($N_P^{MC}$) | 1250 | 10152 | $6.73{\cdot}10^{-2}$ | $4.58{\cdot}10^{-3}$ | $6.81{\cdot}10^{-2}$ | 4.69 |

Comparison against the standard MC simulation. The subscripts *nom* and *pert* refer to the nominal and perturbed solutions, respectively

**Table 8** Application of LS with a perturbed sampling direction to the case of asteroid 99942 Apophis

| | $N_{IC}$ | $N_P$ | $\hat{P}(I)$ | $\hat{\sigma}(P(I))$ | $\delta$ | FoM |
|---|---|---|---|---|---|---|
| MC | 1000000 | 1000000 | $3.00{\cdot}10^{-5}$ | $5.48{\cdot}10^{-6}$ | $1.83{\cdot}10^{-1}$ | $3.33{\cdot}10^{4}$ |
| $LS_{nom}$ (ref) | 1000 | 10351 | $3.12{\cdot}10^{-5}$ | $2.64{\cdot}10^{-7}$ | $8.44{\cdot}10^{-3}$ | $1.39{\cdot}10^{9}$ |
| $LS_{nom}$ ($\sigma^{MC}$) | 216 | 2511 | $3.19{\cdot}10^{-5}$ | $5.48{\cdot}10^{-7}$ | $1.72{\cdot}10^{-2}$ | $1.32{\cdot}10^{9}$ |
| $LS_{nom}$ ($N_P^{MC}$) | 100000 | 1000351 | $3.10{\cdot}10^{-5}$ | $2.56{\cdot}10^{-8}$ | $8.26{\cdot}10^{-4}$ | $1.53{\cdot}10^{9}$ |
| $LS_{pert}$ (ref) | 10000 | 100351 | $2.73{\cdot}10^{-5}$ | $1.31{\cdot}10^{-7}$ | $4.81{\cdot}10^{-3}$ | $5.76{\cdot}10^{8}$ |
| $LS_{pert}$ ($\sigma^{MC}$) | 530 | 5651 | $2.76{\cdot}10^{-5}$ | $5.48{\cdot}10^{-7}$ | $1.99{\cdot}10^{-2}$ | $5.89{\cdot}10^{8}$ |
| $LS_{pert}$ ($N_P^{MC}$) | 100000 | 1000351 | $2.74{\cdot}10^{-5}$ | $4.15{\cdot}10^{-8}$ | $1.52{\cdot}10^{-3}$ | $5.80{\cdot}10^{8}$ |

Comparison against the standard MC simulation. The subscripts *nom* and *pert* refer to the nominal and perturbed solutions, respectively

It is notable, however, how the LS method in both these cases can uncover parts of the impact regions that were not detected via the standard MC sampling or via a different sampling direction. This is particularly visible in the case of 2010 $RF_{12}$, where the sampling using the new direction can uncover a curved impact region extending outside the $3\sigma$ uncertainty distribution, as seen in Fig. 7. In this case, in fact, the green curve represents the real shape of the impact region, which was not found by the standard MC since those solutions are at the boundaries of the initial distribution. In the nominal case presented in Fig. 3, where the sampling direction follows the main axis of the uncertainty distribution, the sampling lines are almost entirely contained inside the $3\sigma$ distribution, and only the areas of the impact region that intersect with it are found. In the new case, however, the sampling direction is directed outside the initial distribution: this allows the sampling lines to intersect the impact region outside the $3\sigma$ distribution. These solutions are several standard deviations away from the nominal solution, thus their contribution to the estimated impact probability is negligible, as the similar numerical results in Table 7 show.

### 5.1.2 Identification of the boundaries

The identification of the boundaries of the impact region is another key parameter of the method. Due to the nonlinear nature of the problem under analysis, they must be determined numerically. The accuracy of this numerical process affects the performance of LS, since any error in $(\bar{c}_1^k, \bar{c}_2^k)$ propagates into the conditional impact probabilities, and in turn into the estimates of the overall probability and its associated standard deviation.

In the implementation adopted for this work, the intersections with the impact region along each sampling line are determined via Newton's iterations and interpolation. The result is the numerical approximation of the values of $c^k$ (the parameter describing the $k$-th sampling line) at the two intersection points, i.e. $(\bar{c}_1^k, \bar{c}_2^k)$. It is worth specifying that, since no information about the derivative of the performance function is available (see Sect. 2), each iteration involves two orbital propagations, which are needed for the numerical differentiation.

To show how the accuracy of this numerical process affects the accuracy of the impact probability estimate, asteroids 2010 $RF_{12}$ and 99942 Apophis are again considered as reference test cases. Starting from the same random initial conditions associated with the nominal results presented in Sect. 4, a lower number of Newton's iterations is here imposed, which worsens the accuracy of the approximation of the intersection points along each sampling

**Table 9** Application of LS with a lower number of Newton's iterations to the case of asteroid 2010 $RF_{12}$

| | $N_{IC}$ | $N_P$ | $\hat{P}(I)$ | $\hat{\sigma}(P(I))$ | $\delta$ | FoM |
|---|---|---|---|---|---|---|
| MC | 10000 | 10000 | $6.51{\cdot}10^{-2}$ | $2.47{\cdot}10^{-3}$ | $3.79{\cdot}10^{-2}$ | 16.43 |
| $LS_{nom}$ (4 iter.) | 1000 | 8245 | $6.57{\cdot}10^{-2}$ | $3.30{\cdot}10^{-3}$ | $5.02{\cdot}10^{-2}$ | 11.22 |
| $LS_{sens}$ (3 iter.) | 1000 | 6245 | $7.25{\cdot}10^{-2}$ | $3.41{\cdot}10^{-3}$ | $4.70{\cdot}10^{-2}$ | 13.93 |
| $LS_{sens}$ (2 iter.) | 1000 | 4245 | $1.35{\cdot}10^{-1}$ | $3.72{\cdot}10^{-2}$ | $2.75{\cdot}10^{-2}$ | 17.32 |

Comparison against the standard MC simulation. The subscripts *nom* and *sens* refer to the nominal solution and the results of the sensitivity analysis, respectively

**Table 10** Application of LS with a lower number of Newton's iterations to the case of asteroid 99942 Apophis. Comparison against the standard MC simulation

| | $N_{IC}$ | $N_P$ | $\hat{P}(I)$ | $\hat{\sigma}(P(I))$ | $\delta$ | FoM |
|---|---|---|---|---|---|---|
| MC | 1000000 | 1000000 | $3.00{\cdot}10^{-5}$ | $5.48{\cdot}10^{-6}$ | $1.83{\cdot}10^{-1}$ | $3.33{\cdot}10^{4}$ |
| $LS_{nom}$ (5 iter.) | 1000 | 10351 | $3.12{\cdot}10^{-5}$ | $2.64{\cdot}10^{-7}$ | $8.44{\cdot}10^{-3}$ | $1.39{\cdot}10^{9}$ |
| $LS_{sens}$ (4 iter.) | 1000 | 8351 | $3.15{\cdot}10^{-5}$ | $1.91{\cdot}10^{-7}$ | $6.05{\cdot}10^{-3}$ | $3.29{\cdot}10^{9}$ |
| $LS_{sens}$ (3 iter.) | 1000 | 6351 | $3.19{\cdot}10^{-5}$ | $1.08{\cdot}10^{-7}$ | $3.38{\cdot}10^{-3}$ | $1.36{\cdot}10^{10}$ |

The subscripts *nom* and *sens* refer to the nominal solution and the results of the sensitivity analysis, respectively line. Tables 9 and 10 show the results of the numerical simulations. The results are reported for the same number of sampling lines and are compared with those of standard MC and with the nominal results shown in Tables 4 and 6. Both analyses show that an incorrect determination of the boundaries of the impact region can lead to inaccurate impact probability estimates, even though comparable values of $\hat{\sigma}(P(I))$ are obtained in both cases (as already explained in Sect. 2.4, this depends on the number of line integrals that are evaluated). This behaviour can be explained by considering that the overall probability is estimated through the sum of 1D integrals, which are computed on each sampling line between the two $(\overline{c}_1^k, \overline{c}_2^k)$ values. Thus, evaluating the integrals with different values is equivalent to considering a different impact region with a different associated impact probability. While using one fewer iteration does not affect the value of the probability in a relevant way, especially in the case of 99942 Apophis, further reductions could yield inaccurate estimates. The minimum number of iterations for a correct estimate depends on the probability level (the lower the probability is, the more difficult is to identify the boundaries, and thus, more iterations are required), on the shape of the impact region, and on the sampling direction itself, and must be selected as a trade-off between accuracy and computational effort.

## 5.2 Subset simulation

The analysis presented in Sect. 4 was carried out considering different values of $N$, whereas predefined $p_0$ and auxiliary distribution were selected. While a general description of the role of the proposal distribution was offered in Sect. 3, no guideline for the selection of the conditional probability was offered. An analysis is here presented.

The conditional probability $p_0$ is a critical parameter of the SS method. It affects the number of intermediate regions $I_j$ that are needed to reach the target impact region, and in turn the efficiency of SS (Zuev et al. 2012). Smaller $p_0$ values grant fewer intermediate levels, though require a larger number of samples per conditional level for the accurate estimation of the small conditional probability. Conversely, the number of samples needed per conditional level can be reduced by increasing the value of $p_0$, with the drawback of increasing the number of conditional levels to reach the final impact region as well. The selection of $p_0$ must take all these considerations into account.

The value of $p_0$ also affects the overall number of samples. Assuming not to alter the value of $N$, the expression for the total number of samples

$$N_T = N + (1 - p_0)(n - 1)N \tag{25}$$

shows that $N_T$ depends on both $p_0$ and $n$.

The aim of our analysis is to investigate the impact of $p_0$ on the performance of the SS method. In order to decouple the effect of $p_0$ and $N_T$, the analysis is carried out by fixing the overall number of samples and changing the value of $p_0$. This of course implies that the value of $N$ must be tuned accordingly. This value is selected by estimating the required number of conditional samples based on the known estimate of the impact probability for the specific test case. Given a test case with estimated impact probability $\hat{P}$, the procedure is the following:

1. Select the value of $p_0$.
2. Select the overall number of samples $N_T$.
3. Guess the number of conditional levels $n$.
4. Compute the value of $N$ from Eq. (25). The number must be an integer value, so round the value to the closest integer.
5. Use Eq. (25) to change the value of $p_0$ with the rounded value of $N$.
6. Compute $N_I$ by inverting Eq. (22) and round the value to the closest integer.
7. Use Eq. (22) to change the value of $\hat{P}$.
8. Consider $N_I$: if $N_I \leq Np_0$, stop the algorithm, otherwise set $n = n + 1$ and go back to step 5.

The above algorithm provides the estimate of the value of $N$ required to start the SS method with different values of $p_0$, without changing the overall number of samples.

The analysis is run on the three test cases of Sect. 4. The overall number of samples is set to 10000, which is sufficiently large to be used in all the test cases. Different values of $p_0$ are tested. The lower limit ($p_0 = 0.001$) was selected to be compatible with the value of $N_T$ in case the simulation performs only one iteration, i.e. in case SS collapses into an MC simulation. The upper limit, instead, was selected very close to 1 (0.99), since the SS procedure is singular for $p_0 = 1$. Figure 9 shows the trend of $\delta$ for all the test cases and for increasing values of $p_0$. As can be seen, the test cases of asteroids 2017 $RH_{16}$ and 99942 Apophis show a general decreasing trend, whereas only small variations appear for asteroid 2010 $RF_{12}$. In this case, $\delta$ initially increases while moving from very low values of $p_0$ (i.e. MC simulation) to larger values. Then, it reaches its minimum at $p_0 \approx 0.4$ and finally slightly increases again. This trend confirms the results reported in Sect. 4 for the same test case, where MC outperformed SS with a value of conditional probability equal to 0.1.

As can be seen from Fig. 9, the trend of the coefficient of variation depends on the impact probability, with the worst performance obtained in general when SS degenerates into MC, and with a decrease in $\delta$ for increasing values of $p_0$. Nevertheless, when the value of conditional probability exceeds 0.2–0.3, no significant improvement can be obtained in terms of accuracy. It should be noted that, if the overall number of samples is fixed, the number of conditional levels increases, and the number of samples per conditional level decreases. In practical situations, when the values of $N$ and $p_0$ are selected a priori and the overall number of samples is automatically determined by the required number of conditional samples, increasing $p_0$ may provide improvements in terms of accuracy, with the drawback of increasing the number of propagated samples and required pointwise propagations. Therefore, it is reasonable to say that selecting $p_0$ out of the range 0.1–0.4 does not necessarily grant sufficient accuracy benefits to the SS method. The obtained results confirm what was presented by Zuev et al. (2012). In their analyses, they show that, assuming not to alter the overall number of samples, the value of conditional probability that minimises the coefficient of variation $\delta$ is approximately 0.2. As a general trend, it is shown that, if one selects a value of $p_0$ between 0.1 and 0.3, similar results in efficiency are obtained. These results are consistent with the analysis shown in this section, and can be used as a general guideline when applying the SS method.

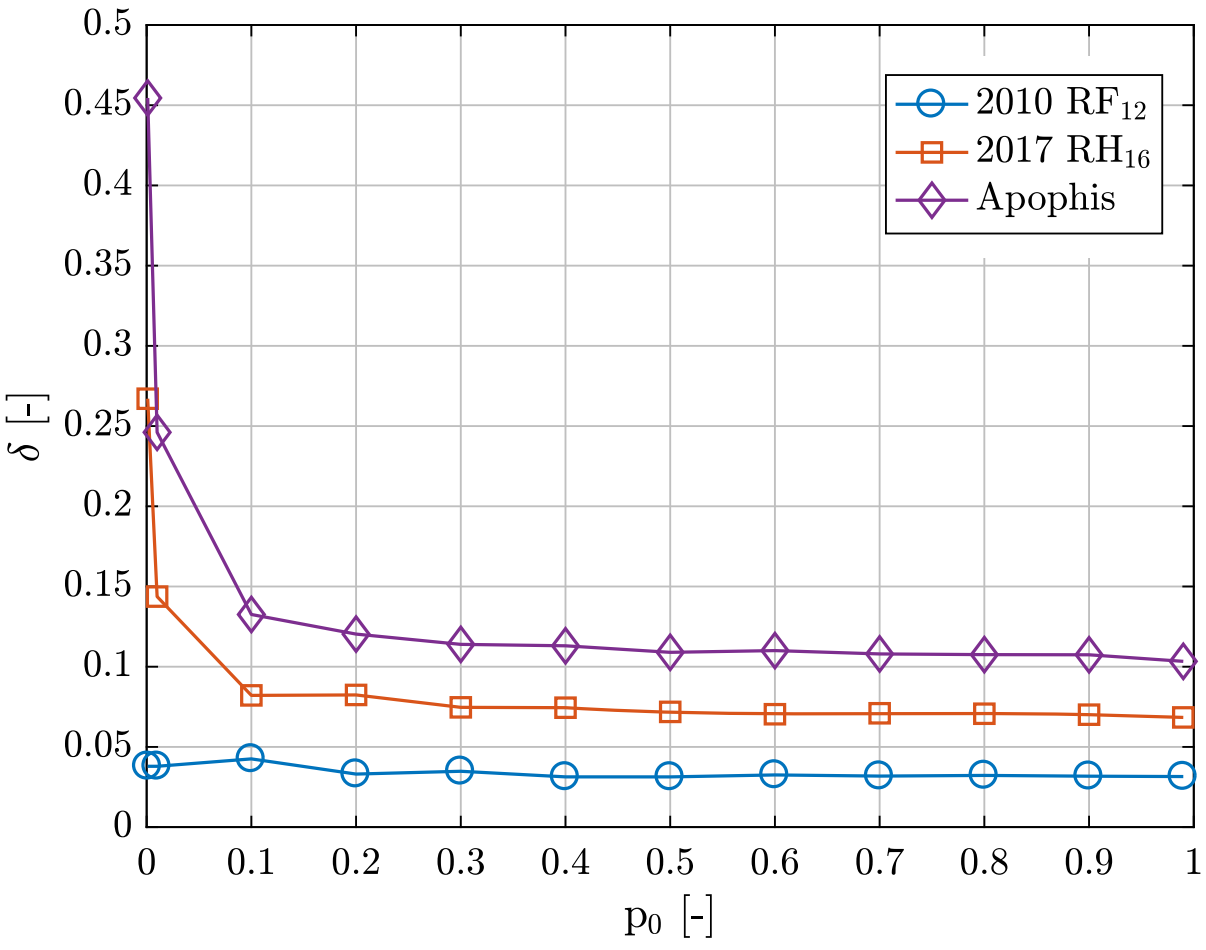

**Fig. 9** SS method: coefficient of variation as a function of the conditional probability level $p_0$ for the three test cases

A final remark on the SS setup procedure is finally needed. In the examples shown in this paper, known test cases have been investigated, so the problem of the identification of the epoch and the number of possible impact regions has not been addressed. In real cases, with no a priori information about the number of potential close approaches, the procedure that should be followed is the one described in Sect. 3. It is evident, however, that if multiple simulations need to be run to characterise all possible impact regions, the advantages granted by the SS with respect to standard MC still exist but are significantly reduced. This is the main reason why, in the last years, new approaches for the characterisation in parallel of multiple failure regions have been proposed (Hsu and Ching 2010; Li et al. 2015). These approaches, which could be very interesting for NEA applications, have not been tested yet by the authors, but are worth studying.

## 6 Conclusions and future work

This paper introduced the application of two advanced Monte Carlo sampling methods, line sampling and subset simulation, to the critical issue of computing the impact probability of near-Earth asteroids. A detailed analysis of both methods was presented in the first part of the paper, offering some guidelines and an illustration of the most critical points of their implementation. Three different test cases were then presented and the performance of the two methods were compared against standard Monte Carlo. Both methods allow significant savings in computational time with respect to standard MC in case of rare and very rare events, as they reduce the number of required propagations while granting the same level of accuracy by either identifying optimal sampling directions or limiting the sampling to specific regions of the phase space. Further analysis has shown that the performance of each method can suffer from particular choices of the parameters characterising their implementation, affecting either the accuracy of the probability estimate or the computational load. Overall, the analyses presented in this paper show that both methods can represent a valuable alternative to standard MC and could be used for impact probability computation every time the operational approach adopted in this field (the LOV method) cannot be applied or grows in complexity. Yet, some improvements are still possible.

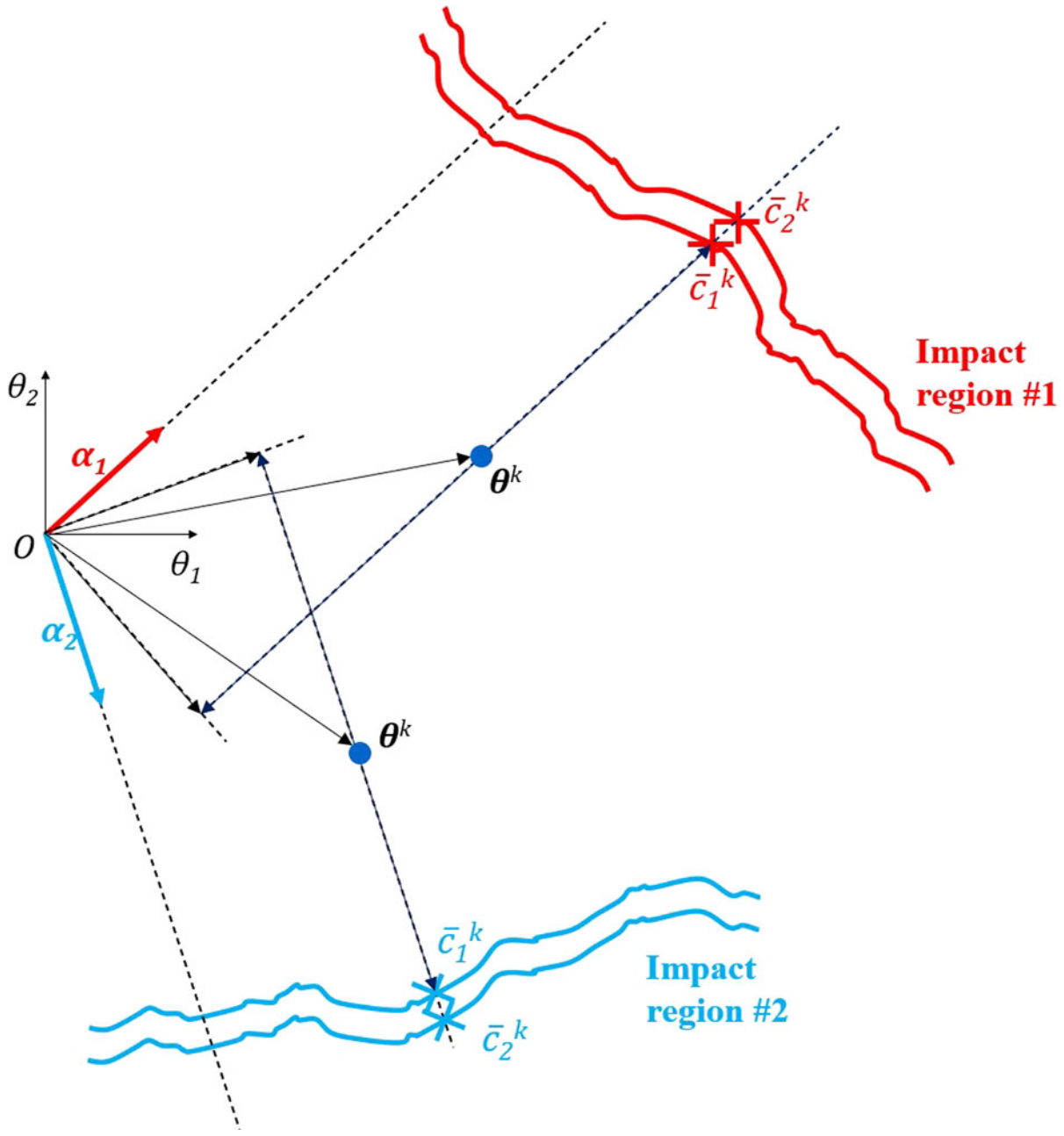

**Fig. 10** Application of the LS method for two impact regions

Future work will focus on improving the efficiency of the two methods. Part of the work will be dedicated to the improvement of the identification of the boundaries of the impact region for the LS method, by finding more efficient iterative procedures. A significant effort will be then devoted to investigate the applicability of the LS method to scenarios with disconnected impact regions over large time spans. In this regard, an improved algorithm based on LS has been recently proposed in Romano (2020) and Romano et al. (2020) and applied to cases where multiple impact regions are expected. In these references, the presented algorithm is able to identify the time intervals where the possible impact regions are located and analyse them via multiple applications of line sampling, by changing the sampling direction to investigate each of them in an optimal way. An example of how this improved algorithm could be applied is shown in Fig. 10, which provides a visual explanation of how this possible enhancement works when applied to the analysis of two impact regions.

In parallel, an analysis on the sensitivity of the SS method to the proposal distribution will be carried out, with the aim of obtaining more rigorous guidelines for its selection. All these activities will be carried out to investigate the possibility of coupling the action of the two methods in a single sampling technique. In addition, the proposed methods will be tested against a wider range of cases and scenarios, and a rigorous comparison with the LOV approach will be performed.

**Acknowledgements** Part of the research leading to these results has received funding from the European Research Council (ERC) under the European Union's Horizon 2020 research and innovation programme as part of project COMPASS (Grant Agreement No. 679086). The data sets generated for this study can be found in the repository at the link https://www.compass.polimi.it/publications/. M. Romano and M. Losacco gratefully acknowledge professors E. Zio, N. Pedroni and F. Cadini from Politecnico di Milano for their introduction to the Line Sampling and Subset Simulation methods.

**Funding** Open access funding provided by Politecnico di Milano within the CRUI-CARE Agreement.

## Compliance with ethical standards

**Conflict of interest** On behalf of all authors, the corresponding author states that there is no conflict of interest.